\documentclass[aps,a4paper,showkeys,nofootinbib,longbibliography,notitlepage,twocolumn]{revtex4-2}
\usepackage{times}
\usepackage{ulem}
\usepackage{comment}
\usepackage{graphicx}
\usepackage{feynmf}
\usepackage{array}
\usepackage{tabularx}
\usepackage{amsmath}
\usepackage{amstext}
\usepackage{amssymb}
\usepackage{pifont}
\usepackage{xfrac}
\usepackage[colorlinks,citecolor=blue]{hyperref}
\usepackage{graphicx}
\usepackage{amsmath}
\usepackage{amstext}
\usepackage{amssymb}
\usepackage{amsfonts}
\usepackage{longtable,booktabs}
\usepackage{hyperref}
\usepackage{url}
\usepackage{subfigure}
\usepackage{dsfont}
\usepackage{booktabs}
\usepackage{amsbsy}
\usepackage{dcolumn}
\usepackage{amsthm}
\usepackage{bm}
\usepackage{esint}
\usepackage{multirow}
\usepackage{hyperref}
\usepackage{cleveref}
\usepackage{mathrsfs}
\usepackage{amsfonts}
\usepackage{amsbsy}
\usepackage{dcolumn}
\usepackage{bm}
\usepackage{multirow}
\usepackage{color}
\usepackage{extarrows}
\usepackage{datetime}
\usepackage{orcidlink}
\usepackage[super]{nth}
\hypersetup{
	colorlinks=magenta,
	linkcolor=blue,
	filecolor=magenta,
	urlcolor=magenta,
}

\newcommand{\Tr}{\text{Tr}}

\begin{document}
\title{Emergent Non-Markovian Gain in Open Quantum Systems}

\author{H. Z. Shen~\orcidlink{0000-0002-4017-7367}$^{1}$}
\email{Contact author: shenhz458@nenu.edu.cn}

\author{Cheng Shang~\orcidlink{0000-0001-8393-2329}$^{2,3}$}
\email{Contact author: cheng.shang@riken.jp}

\author{Yan-Hui Zhou$^{4}$}
\email{Contact author: yanhuizhou@126.com}

\author{X. X. Yi$^{1}$}
\email{Contact author: yixx@nenu.ed.cn}

\makeatletter
\renewcommand\frontmatter@affiliationfont{\vspace{1mm} \small}
\makeatother

\affiliation{\textit{$^{1}$Center for Quantum Sciences and School of Physics, Northeast Normal University, Changchun 130024, China\\$^{2}$Analytical quantum complexity RIKEN Hakubi Research Team, RIKEN Center for Quantum Computing (RQC), Wako, Saitama 351-0198, Japan\\$^{3}$Department of Physics, The University of Tokyo, 5-1-5 Kashiwanoha, Kashiwa, Chiba 277-8574, Japan\\$^{4}$Quantum Information Research Center and Jiangxi Province Key Laboratory of Applied Optical Technology, Shangrao Normal University, Shangrao 334001, China}}

\date{\today}

\begin{abstract}
    Non-Markovian dynamics go beyond the Markovian approximation by capturing memory effects and information backflow in open quantum systems, which are crucial for describing realistic physical processes. In this work, we study the exact non-Markovian dynamics of a driven cavity coupled to an anisotropic three-dimensional photonic-crystal environment via counterrotating-wave interactions. We derive an exact analytical expression for the cavity amplitude satisfying the integro-differential equation, which includes the contributions of the bound states outside the continuum and the dissipative parts with the continuum spectrum. Based on the characteristic function method, we derive the exact non-Markovian master equation for the cavity, which contributes to the gain of the cavity. We give the physical origin of non-Markovian gain in the presence of bound states in the system consisting of cavity and environment, which has no Markovian counterparts due to the nonexponential gain in the non-Markovian structured environment. We find that three different types of bound states can be formed in the system, containing one bound state with no inversion of photon number, two bound states with the periodic equal-amplitude oscillation, and the gain with two complex roots without the bound states formation. We derive a current equation including the source from the driving field, the transient current induced by the change in the number of photons, and the two-photon current caused by the counterrotating-wave term. The results are compared with those given by the rotating-wave interactions and extended to a more general quantum network involving an arbitrary number of coupled cavities. Our findings may pave the way for a deeper understanding of non-Markovian dynamics with gain in quantum networks involving counterrotating-wave effects.
\end{abstract}
\maketitle
\section{Introduction}\label{section 1}
The dynamics of open quantum systems is a subject of ongoing research and growing interest. This is due to its pivotal role in the theoretical framework of quantum physics and the remarkable advancements in quantum technologies. These technological leaps have brought the study of open quantum system dynamics into the spotlight, making it an area of intense focus, with extensive research into the precise characterization of open quantum systems. A significant portion of this research centers around two main approaches. One approach is centered on quantum Brownian motion, which is based on the Feynman-Vernon influence functional~\cite{Caldeira1215871983,PhysRevA.32.2462,Feynman1963TheTO}. The other approach focuses on the stochastic diffusion Schr\"{o}dinger equation~\cite{PhysRevA.58.1699,PhysRevLett.83.4909,PhysRevA.69.052115}. More refined methods for dealing with the system-environment strong-coupling regime have been proposed, such as the Hu-Paz-Zhang~\cite{PhysRevD.45.2843} master equation with time-dependent coefficients, which enables non-Markovian dynamics. Notably, Halliwell and Yu~\cite{PhysRevD.53.2012} have presented an alternative derivation of the Hu-Paz-Zhang equation, in which the dynamics are represented by the Wigner function. Moreover, Ford and O'Connell~\cite{PhysRevD.64.105020} provided an exact solution to this equation. Based on the von Neumann approach to reducing the state vector~\cite{vonNeumann1955}, these contributions were primarily driven by the pioneering efforts of Zurek~\cite{PhysRevD.24.1516,PhysRevD.26.1862}, Caldeira, Leggett~\cite{Caldeira_1983,PhysRevA.31.1059}, Joos, and Zeh~\cite{Joos1985TheEO}. Their work has influenced the field, sparking renewed interest in open quantum systems in various nanostructures~\cite{PhysRevB.78.235311,Tu2009ExactME,Jin2010NonequilibriumQT}, the master equation for microcavities or nanocavities in photonic crystals, and the quantum transport theory for photonic crystals~\cite{PhysRevLett.77.3272,PhysRevA.82.012105,Wu2010NonMarkovianDO,li2025macroscopic}.

The dissipation quantum dynamics of optical cavities have been well investigated and deeply understood under the Markovian approximation~\cite{Carmichael1993,Breuer2002,Gardiner2004,Weiss2008,RevModPhys.88.021002}, which is valid when the coupling between the system and the environment is weak enough to apply the perturbation. In the meantime, the characteristic time of the environment is sufficiently shorter than that of the system so that the non-Markovian memory effect is negligible. However, in many situations in the recent development of optical microcavities, the strong coupling or long-time memory effect has become an important factor in controlling cavity dynamics. Typical examples include optical fields propagating in cavity arrays or optical fiber~\cite{PhysRevLett.79.5242,Hartmann2006StronglyIP,PhysRevA.74.062303,Shang2018NonreciprocityIA,shang2023coupling,luan2025nonreciprocal,zhang2025manipulating}, trapped ions subjected to artificial colored noise~\cite{PhysRevA.62.053807,Myatt2000DecoherenceOQ,PhysRevA.69.052101,PhysRevA.74.032303,PhysRevA.79.052120}, microcavities interacting with a coupled resonator optical waveguide (CROW) or photonic crystals~\cite{PhysRevB.57.12127,Yariv1999CoupledresonatorOW,PhysRevLett.84.2140,PhysRevE.62.7389,Olivier2001MinibandTI,PhysRevB.72.165330,Liu2005ModelockingOM,PhysRevA.74.063826,Longhi2006BoundSI,PhysRevA.62.013805,PhysRevLett.84.2136,PhysRevApplied.21.044048}, and so on. Specifically, for the trapped ions coupled with an engineered environment, the change of the characteristic frequency of the environment can be accomplished simply by applying a random electric field through a band-pass filter defining the frequency spectrum of the environment~\cite{PhysRevA.62.053807,Myatt2000DecoherenceOQ}. On the other hand, for a cavity interacting with CROW or photonic crystals, the coupling between them is controllable by changing the geometrical parameters of the defect cavity and the distance between the cavity and the CROW~\cite{Liu2005ModelockingOM}. Both of them provide non-Markovian dissipation and decoherence channels~\cite{PhysRevA.69.052101,PhysRevA.74.032303,PhysRevA.79.052120,PhysRevA.74.063826}. These strong coupling or long-time memory effects result in a complicated non-Markovian process in cavity systems that have become a crucial concern for the rapid development of quantum information and quantum computation in terms of photons~\cite{_Nielsen_Chuang_2010}. The non-Markovian behavior of the trapped ions has been discussed in many works~\cite{PhysRevA.69.062107,PhysRevA.79.032102,PhysRevA.81.052105,PhysRevA.83.032102,PhysRevA.88.033835,PhysRevE.93.012107,PhysRevA.96.033805,PhysRevE.95.012156}.

Moreover, the excitation backflow effects have been involved in multiple environments feedback on systems~\cite{Gardiner2004,PhysRevB.70.045323,PhysRevA.80.042112,PhysRevA.81.062124,PhysRevA.81.042103,Weiss2008,Franco2012DYNAMICSOQ,PhysRevA.89.042117,PhysRevA.92.012315,RevModPhys.89.015001,PhysRevA.107.053705}, quantum feedback control~\cite{PhysRevA.86.052304}, quantum channel capacity~\cite{Bylicka2014NonMarkovianityAR}, coupled cavities~\cite{PRXQuantum.3.020348}, photonic crystals~\cite{Noda2007SpontaneousemissionCB,Liu2016QuantumEN}, color noises~\cite{PhysRevA.95.052126}, and the cavity and atom coupled to waveguides~\cite{PhysRevA.105.053706,lu2024topologicalquantumbatteries}. These non-Markovian systems have been experimentally realized~\cite{PhysRevA.47.3202,PhysRevA.82.042328,PhysRevLett.106.233601,Liu2011ExperimentalCO,Tang2011MeasuringNO,PhysRevLett.112.210402,Haseli_2014,Grblacher2013ObservationON,PhysRevA.100.052104,PhysRevA.99.022107,Uriri_2020,PhysRevA.102.062208,PhysRevLett.126.230401,PhysRevA.104.022432,PhysRevLett.129.140501,PhysRevLett.128.200601}. The excitation backflow between systems and their environments can be used to characterize the non-Markovian effects of the environment on the system dynamics~\cite{RevModPhys.88.021002,PhysRevLett.103.210401,PhysRevA.81.062115,PhysRevA.81.044105,PhysRevA.83.012108,PhysRevA.86.062108,PhysRevA.90.052103,PhysRevA.92.042108}, and there are various measures of non-Markovianity~\cite{PhysRevLett.101.150402,PhysRevLett.105.050403,PhysRevA.82.042103,PhysRevA.83.062115,PhysRevA.86.044101,PhysRevA.86.012101,PhysRevA.88.020102,PhysRevLett.112.120404}.

The rotating-wave approximation (RWA) is widely used in quantum optics, which neglects the rapidly oscillating counter-rotating terms, where the system Hamiltonian becomes time-independent or slightly time-dependent in the rotating frame. With recent developments in the area of circuit and cavity QED systems~\cite{Wallraff2004StrongCO,Niemczyk2010CircuitQE,PhysRevLett.105.237001}, ultra-strong and deep-strong light-matter couplings have become experimentally achievable. This makes it necessary to take the counter-rotating terms into account. Recent studies show that the counter-rotating terms in system-environment coupling play important roles in non-Markovian effects.

Previous studies of the exact non-Markovian master equation~\cite{PhysRev.36.823,PhysRevE.54.2084,PhysRevLett.105.240403} for cavities are mainly based on two methods. One is that the system environment in the strong coupling regime is in the absence of an external driving field. This method includes the characteristic function~\cite{PhysRevA.90.032105}, the adjoint master equation based on the Heisenberg picture~\cite{PhysRevA.95.052119}, the Lindblad master equation~\cite{PhysRevA.94.042123}, the momentum coupling model~\cite{PhysRevA.96.012109}, exact master equations~\cite{PhysRevE.68.026111}, and the Heisenberg-Langevin equation~\cite{PhysRevA.96.062108}. The other is that of the driving system-environment coupling under the rotating-wave approximation~\cite{PhysRevA.97.042121}. Since non-Markovian behavior is sensitive to the counter-rotating terms in the interaction Hamiltonian~\cite{PhysRevLett.113.200403,PhysRevLett.116.120402,PhysRevA.98.023856,PhysRevA.101.013826,PhysRevA.109.043714,PhysRevLett.132.170402}, important dynamics features can be omitted under the RWA in the strong coupling regime.

Nevertheless, several questions inevitably emerge: (i) How can one get a driving cavity coupling with an environment via counter-rotating-wave interactions? (ii) Is it feasible to extend the driving cavity coupling with an environment through counter-rotating-wave interactions from Markovian systems to non-Markovian ones? (iii) Does the driving field affect the time-dependent coefficients (free term and dissipation) in the exact non-Markovian master equation for the system?

In response to these queries, we propose a scheme to study the exact non-Markovian dynamics with counter-rotating-wave interactions between a driving cavity and an anisotropic three-dimensional photonic crystal environment. We obtain an analytical solution for the cavity amplitude, which includes the contributions from both the bound state part outside the continuum and the dissipative part with the continuum spectrum. By relying on the characteristic function method, we derive the exact non-Markovian master equation of the cavity, which leads to the gain of the cavity. We give the physical origin of this gain when bound states exist in the system composed of the cavity and the environment. We discover three distinct types of bound states that can be formed within the system: static bound states without dynamical inversion, periodic equal-amplitude oscillations with two bound states, and gain with two complex roots. We derive a current equation that consists of the source from the driving field, the transient current flowing from the system into the environment, and the two-photon current induced by the counter-rotating-wave term. Subsequently, the results we present are contrasted with those obtained by the rotating-wave interactions.

The remainder of this paper is organized as follows. In Sec.~\ref{section II}, we derive the non-Markovian Heisenberg-Langevin equation for the cavity operator with the driving field. In Sec.~\ref{section III}, we analytically give the calculation of the cavity amplitude. Sec.~~\ref{section IV} is dedicated to the non-Markovian master equation of the driving cavity. In Sec.~\ref{section V}, we study bound state and non-Markovian dynamics. In Sec.~\ref{section VI}, we investigate the relationship between the reduced density matrix of the driving and non-driving cavities. In Sec.~\ref{section VII}, we study transient current. Finally, we conclude in Sec.~\ref{section VIII}.

\section{Non-Markovian Heisenberg-Langevin equation for cavity operator}\label{section II}
The Hamiltonian of the counterrotating-wave interactions between the driving cavity and the structured environment in the rotating frame with the driving frequency ${{\omega _l}}$ can be written as
\begin{align}
\hat H(t) = \ &   \Delta {\hat a^\dag }\hat a + \sum\limits_k {{{\tilde \omega }_k}} \hat b_k^\dag {\hat b_k} + i\sum\limits_k {(g_k^*{{\hat a}^\dag }\hat b_k^\dag  - {g_k}\hat a{{\hat b}_k})} \nonumber\\ &\!+ \ F(t){\hat a^\dag } + {F^*}(t)\hat a,
\label{eq2}
\end{align}
where $\Delta=\omega_a-\omega_l$ and ${\tilde {\omega} _k}=\omega _k+\omega _l$. Here, ${\hat a}^\dag$ and $\hat a$ are the creation and annihilation operators of the cavity field, whose frequency is $\omega_a$. Additionally, ${\hat b}^\dag$ and $\hat b$ are the collections of infinite harmonic oscillators creation and annihilation operators of the $k$-th oscillator with the frequency $\omega_k$. The third term describes the interaction between the cavity and the environment with the coupling strength $g_k$. The last two terms denote the single-photon driving field to the cavity with amplitude $F(t)$ and frequency $\omega _l$. The experimental implementation scheme without the driving field in Eq.~(\ref{eq2}) can be found in Appendix~\ref{section A}. The system and environment operators obey the Heisenberg equation
\begin{eqnarray}
\frac{d}{{dt}}\hat a(t) &=&  - i{\Delta}\hat a (t)+ \sum\limits_k {g_k^*} \hat b_k^\dag (t) - iF(t),\label{eq3}\\
\frac{d}{{dt}}{{\hat b}_k}(t) &=&   - i{\omega_k}{\hat b_k}(t) + g_k^*{\hat a^\dag (t)}.
\label{eq4}
\end{eqnarray}
Substituting solution of Eq.~(\ref{eq4}) into Eq.~(\ref{eq3}), we obtain
\begin{eqnarray}
\begin{aligned}
\!\!\!\!\!\!\frac{d}{{dt}}\hat a(t) \!=\!  - i{\Delta}\hat a(t) \!-\! i\hat B(t) \!+\! \int_0^t \hat a (\tau )g(t - \tau )d\tau  \!-\! iF(t),
\label{eq10}
\end{aligned}
\end{eqnarray}
where
\begin{eqnarray}
\begin{aligned}
g(t) = \int {J(\omega )} {e^{i\omega t}}d\omega,
\label{eq8}
\end{aligned}
\end{eqnarray} 
denotes the correlation function originating from the counter-rotating wave interactions, which is fundamentally different from that of the rotating-wave system represented by
\begin{eqnarray}
\begin{aligned}
{g_{\rm{RWA}}}(t) = \int {J(\omega )} {e^{ - i\omega t}}d\omega.
\label{houjiaeq8}
\end{aligned}
\end{eqnarray}$J(\omega ) = \sum\nolimits_k {{{| {{g_k}} |}^2}} \delta (\omega  - {\omega _k})$ in Eq.~(\ref{eq8}) is the spectral density of the environment, which characterizes all the back-actions between the cavity and the structured environment and can be determined uniquely by the coupled strength ${{| {{g_k}} |}^2}$ through the correlation function (\ref{eq8}). The external environment operator in Eq.~(\ref{eq10}) is ${\hat {B}}(t) = i\sum_k {g_k^*} \hat b_k^\dag {e^{i{\omega_k}t}}$.

Considering the linearity of Hamiltonian~(\ref{eq2}), the cavity operator $\hat a(t)$ can be expressed in terms of the initial components as
\begin{eqnarray}
\begin{aligned}
\hat a(t) = {\cal U}(t)\hat a + \hat f_1(t)+ f_2(t),
\label{eq11c}
\end{aligned}
\end{eqnarray}
where the cavity amplitude satisfies
\begin{eqnarray}
\begin{aligned}
\frac{d}{{dt}}{\cal U}(t) =  - i{\Delta}{\cal U}(t) + \int_0^t {\cal U} (\tau )g(t - \tau )d\tau,
\label{eq14}
\end{aligned}
\end{eqnarray}with environment and driving field parts
\begin{align}
\hat f_1(t) &=  - i\int_0^t {\hat B(\tau )} {\cal U}(t - \tau )d\tau,\nonumber\\
 f_2(t)&= - i\int_0^t {F(\tau )} {\cal U}(t - \tau )d\tau.
\label{eq21}
\end{align}
The integro-differential equation in Eq.~(\ref{eq14}) determines the exact non-Markovian dynamics of the cavity from the structured environment, which constitutes the non-Markovian gain with Eq.~(\ref{eq2}) in counterrotating-wave interactions due to the plus sign in front of the integral in Eq.~(\ref{eq14}), while the minus sign (simultaneously $g(t)$ becomes ${g_{RWA}}(t)$ in Eq.~(\ref{houjiaeq8})) represents the non-Markovian dissipation with the model under the rotating-wave interactions~\cite{Breuer2002,Gardiner2004,Weiss2008,RevModPhys.88.021002}. In the Markovian approximation, the spectral density takes $J(\omega )=\Gamma /{{2\pi }}$ (${\Gamma }$ denotes the dissipation strength), which leads to $g(t)={\Gamma } \delta (t)$. Consequently, Eq.~(\ref{eq14}) is reduced to
\begin{eqnarray}
{\cal U}(t) ={e^{ - i\Delta t + \frac{\Gamma }{{2  }} t}},
\label{jiage}
\end{eqnarray}
which is the pure exponential gain of the cavity induced by the environment with counterrotating-wave interactions.

\section{The calculation of the collective amplitudes ${\cal U}(t) $}\label{section III}
\begin{figure}[htbp]
\centering{
\includegraphics[width=7.2cm, height=8.0cm, clip]{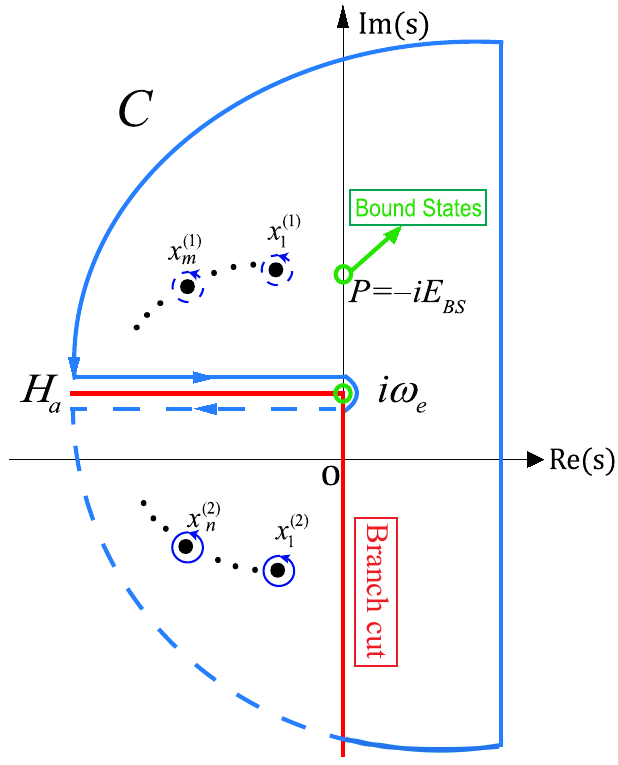}}
\caption{Contour of the inverse Laplace transform of $\tilde {\cal U}(s)$ in Eq.~(\ref{EX-fs1}) with the counterrotating-wave interaction system described by Eq.~(\ref{eq2}). The (red) lower part below point $i{\omega _e}$ of the line on the imaginary axis is the branch cut, where $i \omega_e$ lies in the pure imaginary pole. The integration along the solid (dashed) curve is imposed in the first (second) Riemannian sheet. ${x_m^{(1)}}$ and ${x_m^{(2)}}$ correspond to the poles of $\tilde {\cal U}(s)$ on the first (${\mathop{\rm Re}\nolimits} (s) > 0$ or ${\mathop{\rm Im}\nolimits} (s) > {\omega _e}$) and second (${\mathop{\rm Re}\nolimits} (s) < 0$ and ${\mathop{\rm Im}\nolimits} (s) < {\omega _e}$) Riemannian sheets.}
\label{contour_g}
\end{figure}

The amplitude ${\cal U}(t)$ in Eq.~(\ref{eq14}) can be obtained by means of the inverse Laplace transform
\begin{equation}
\begin{aligned}
{\cal U}(t) = \frac{1}{{2\pi i}}\int_{\sigma  - i\infty }^{\sigma  + i\infty } {\tilde {\cal U}(s){e^{st}}ds},
\label{eatinverse}
\end{aligned}
\end{equation}
where $\tilde {\cal U}(s)$ is given by
\begin{eqnarray}
\begin{aligned}
\tilde {\cal U}(s) = \frac{{1}}{{s + i{\Delta} -g(s)}},
\label{EX-fs1}
\end{aligned}
\end{eqnarray}
with
\begin{eqnarray}
\begin{aligned}
g(s) = \int {\frac{{J(\omega )}}{{s - i\omega }}} d\omega .
\label{EX-fs}
\end{aligned}
\end{eqnarray}
Herein, we take the spectral density of the anisotropic three-dimensional photonic crystal environment as
\begin{eqnarray}
J(\omega ) = \frac{\Gamma}{\pi }\frac{{\sqrt {\omega  - {\omega _e}} }}{\omega }\phi(\omega-\omega_e),
\label{zwjw}
\end{eqnarray}
whose derivations can be found in Appendix~\ref{section B}. Substituting Eq.~(\ref{zwjw}) into Eq.~(\ref{EX-fs}), we obtain frequency-domain correlation function
\begin{eqnarray}
g(s) = \frac{{ i\Gamma}}{{\sqrt {{\omega _{e}}}  + \sqrt {  is + {\omega _{e}}}}},
\label{FS}
\end{eqnarray}
where the phase angle of $s$ is defined by
$- \pi  < \arg (s) < \pi $, the phase angle of $\sqrt {  is + {\omega _{e}}} $ in $g(s)$ is defined
by $ - {\pi  \mathord{\left/
 {\vphantom {\pi  2}} \right.
 \kern-\nulldelimiterspace} 2} < \arg \sqrt {is + {\omega _e}}  < {\pi  \mathord{\left/
 {\vphantom {\pi  2}} \right.
 \kern-\nulldelimiterspace} 2}$. With the integration contours $C$ as shown in Fig.~\ref{contour_g}, we have
\begin{equation}
\begin{aligned}
{\cal U}(t)  = &\sum\limits_m {\frac{{{\cal U}(0){e^{x_m^{(1)}t}}}}{{{{\cal G}^\prime }(x_m^{(1)})}}}  - \frac{{\cal U}(0)}{{2\pi i}} \left\{ {\int_{ i{\omega _e} - \infty }^{ i{\omega _e} + 0} {} } \right. \\
 &\left. { + \int_{  i{\omega _e} + 0}^{  -i\infty  + 0} {ds{e^{st}}\frac{{C(0)}}{{s + i{\Delta} -g(s)}}} } \right\} ,
\label{eatinverse1}
\end{aligned}
\end{equation}
where the function
\begin{equation}
\begin{aligned}
{\cal G}(s) = s + i{\Delta} -g(s),
\label{appGx}
\end{aligned}
\end{equation}
where  ${x_m^{(1)}}$ is the root of the equation ${\cal G}(s)=0$ in the region [${\mathop{\rm Re}\nolimits} (s)>0$ or ${\mathop{\rm Im}\nolimits} (s)>\omega_e$],
the real number $\sigma$, the real number $s=\sigma$ lies to the right of all the singularities ${x_m^{(1)}}$. As shown in Fig.~\ref{contour_g}, the last term in Eq.~(\ref{eatinverse1}) can also be calculated using the integration contours indicated by the dashed lines, as follows:
\begin{equation}
\begin{aligned}
  \!\!- \!\!\sum\limits_n {\frac{{{u(0)e^{x_n^{(2)}t}}}}{{{\cal L}'(x_n^{(2)})}}}  \!-\! \frac{u(0)}{{2\pi i}} \left[ {\int_{i{\omega _e} - \infty }^{  i{\omega _e} + 0} {\!\!\!\!\!ds{e^{st}} \frac{{u(0)}}{{s + i{\Delta} -g_1(s)}}  } } \right],\!\!
\label{applastterm}
\end{aligned}
\end{equation}
where
\begin{eqnarray}
&&{{\cal L}(s) = s + i{\Delta} -g_1(s)}, \label{1appLx} \\
&&{g_1(s) = \frac{{ i\Gamma}}{{\sqrt {{\omega _{e}}}  - \sqrt {  is +{\omega _{e}}} }}},
\label{appLx}
\end{eqnarray}
where ${x_n^{(2)}}$ is the root of the equation ${\cal L}(s)=0$ in the region [${\mathop{\rm Re}\nolimits} (s)<0$ and ${\mathop{\rm Im}\nolimits} (s)<\omega_e$]. Since the closure crosses the branch cut ${\mathop{\rm Im}\nolimits} (s)<\omega_e$ on the imaginary axis, the contour is necessary to pass into the second Riemannian sheet in the section of the half-plane with ${\mathop{\rm Im}\nolimits} (s)<\omega_e$, where it remains in the first Riemannian sheet in the sections ${\mathop{\rm Im}\nolimits} (s)<\omega_e$ in the half plane ${\mathop{\rm Re}\nolimits} (s)<0$ in Fig.~\ref{contour_g}. From Eqs.~(\ref{eatinverse}), (\ref{eatinverse1}), and (\ref{applastterm}), we can obtain the analytical solution of Eq.~(\ref{eq14}) by setting $s=-y+i\omega_e$ on the first (second) Riemannian sheet
\begin{figure}[t]
\centerline{
\includegraphics[width=8.4cm, height=8.0cm, clip]{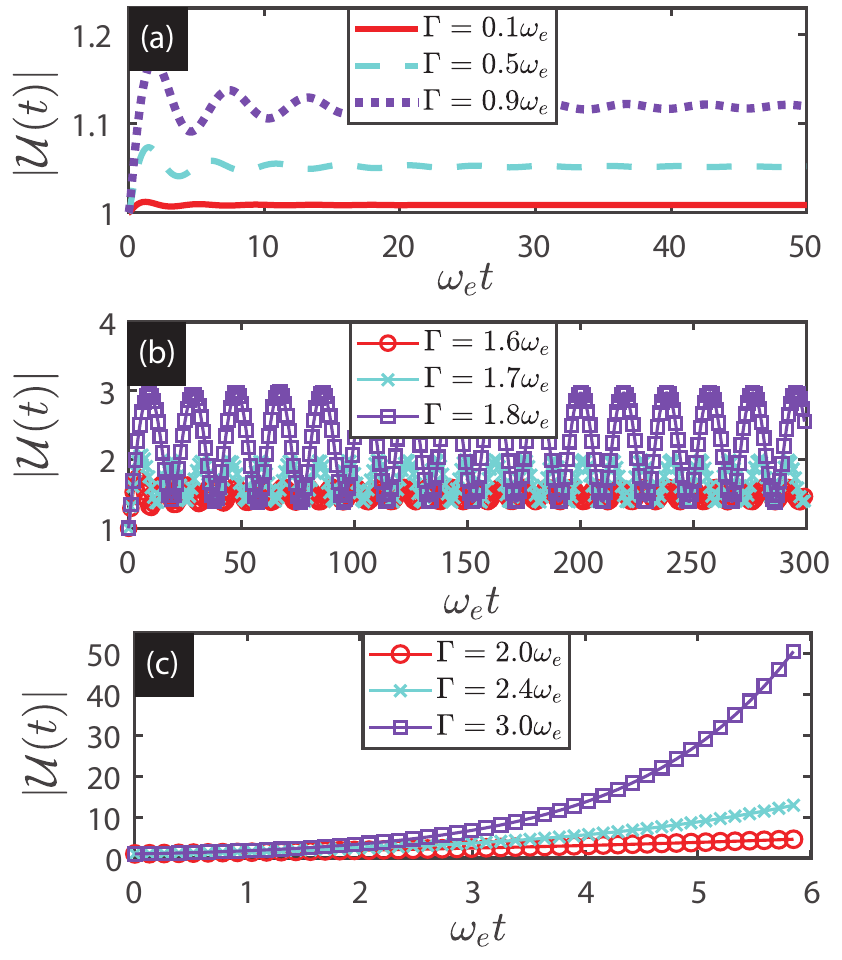}}
\caption{ Time evolution of $|{\cal U}(t)|$ in Eq.~(\ref{calCt}) for the cavity coupled to the non-Markovian environment with the three-dimensional photonic crystal spectrum. This figure corresponds to no inversion, periodic oscillation, and gain, respectively. The parameters are $\Delta = 0.5\omega_e$. For comparison, the red lines denote $\Gamma = 0.1\omega_e$, $\Gamma = 1.6\omega_e$, $\Gamma = 2.0\omega_e$; The blue lines denote $\Gamma = 0.5\omega_e$, $\Gamma = 1.7\omega_e$, $\Gamma = 2.4\omega_e$; The purple lines denote $\Gamma = 0.9\omega_e$, $\Gamma = 1.8\omega_e$, $\Gamma = 3.0\omega_e$.} \label{Ut}
\end{figure}
\begin{eqnarray}
\!\!\!\!{\cal U}(t) \!=\! {\sum\limits_m {\frac{{{e^{x_m^{(1)}t}}}}{{{{\cal G}^\prime }(x_m^{(1)})}}}  \!+\! \sum\limits_n {\frac{{{e^{x_n^{(2)}t}}}}{{{{\cal L}^\prime }(x_n^{(2)})}}} }  { + \frac{1}{{2\pi i}}\int_0^\infty  \!\!\!\!\!{dy} \mu (s){e^{st}}}, \ \ \ \
\label{calCt}
\end{eqnarray}
where $\mu (s) = {{\cal L}^{ - 1}}(s) - {{\cal G}^{ - 1}}(s)$. ${\cal G}'(s)$ and ${\cal L}'(s)$ are derivatives of functions ${\cal G}(s)$ and ${\cal L}(s)$, respectively. $x_m^{(1)}$ and $x_n^{(2)}$ are the roots of ${\cal G}(s)=0$ and ${\cal L}(s)=0$, respectively. The last term in Eq.~(\ref{calCt}) represents the contribution from the contour along the Hankel path $H_a$ in Fig.~\ref{contour_g}. This contribution is accountable for the non-exponential decay dynamics as elaborated in the reference~\cite{CohenTannoudji2008}. Consequently, as demonstrated by Eq.~(\ref{calCt}), the cavity amplitude exhibits dissipationless dynamics on account of the presence of the bound state. This implies that the decoherence of the system can be mitigated by means of strong non-Markovian coupling to an environment. The cavity amplitude ${\cal U}(t)$ in Eq.~(\ref{calCt}) denotes the non-Markovian cavity amplitude with bound state gains, which is different from those in Refs.~\cite{PhysRevLett.133.050401,PhysRevApplied.17.034073,PhysRevA.106.062438,PhysRevA.104.042609,PhysRevLett.109.170402,PhysRevA.93.033833,PhysRevX.6.021027}, where the bound states with dissipations are formed.

Figure~\ref{Ut} shows the relationship between $|{\cal U}(t)|$ and time in the non-Markovian structured environment, which we find that as the value of $\Gamma$ increases, the value of $|{\cal U}(t)|$ also increases. For cavity amplitude with different transition frequencies, we summarize three independent regimes as follows:

(i) {\it No dynamical inversion with gain}. Figure~\ref{Ut}(a) shows that when there is only one bound state (corresponding to a pure imaginary root of Eq.~(\ref{appGx})), the lines first oscillate and eventually tend to stabilize the remaining gain. In this case, there is only one real energy spectrum; therefore, the photon number Eq.~(\ref{calCt}) can be obtained [see Figs.~\ref{Ut}(a)] after a long time
\begin{eqnarray}
\begin{aligned}
|{\cal U}(t \to \infty)| ^2={\left| {\frac{1}{{{{\cal G}^\prime }(x_m^{(1)})}}} \right|},
\label{steadypapbCD2}
\end{aligned}
\end{eqnarray}where ${E}_{\rm{BS}}$ is a pure imaginary root of ${\cal G}(s=x_m^{(1)}=-i{E}_{\rm{BS}})$, while the integral part in Eq.~(\ref{calCt}) approaches zero originating from the Lebesgue-Riemann Lemma~\cite{Bochner1950}. We demonstrate that the photon number holds a nonzero steady value after a long time. This is also understandable because the bound state with gain, as a stationary state of the whole system, has a vanishing decay rate, and the coherence contained in it would be captured during the time evolution.

(ii) {\it Periodic oscillation gain of cavity amplitude with two bound state frequencies}. Interestingly, in this case, quantum interference between the two localized modes at long times $t$ gives rise to periodic oscillations in the dynamics. Figure~\ref{Ut}(b) is a case where two bound states and equal-amplitude oscillations occur. This completely differs from that under the rotating wave approximation, where only one bound state exists in Refs.~\cite{PhysRevLett.133.050401,PhysRevApplied.17.034073,PhysRevA.106.062438,PhysRevA.104.042609,PhysRevLett.109.170402,PhysRevA.93.033833,PhysRevX.6.021027}. This originates from counterrotating-wave interactions between the cavity and environment. The dynamical properties in Fig.~\ref{Ut}(a) and (b) reveal the non-Markovian effects caused by the structured environment, which have no Markovian counterparts due to the pure gain in Eq.~(\ref{jiage}) under the Markovian approximation. The amplitudes of periodic oscillations do not decrease in time. From Eq.~(\ref{calCt}), we obtain the photon numbers for cavity amplitude in the long-time regime
\begin{eqnarray}
\!\!\!\!\!\!\!\!\!\! |{\cal U}(t \to \infty)| \!=\! \sqrt{\sigma_1^2\sigma _2^2 + 2{\sigma _1}{\sigma_2}\cos [({{E}_{\rm{BS}1}} - {{E}_{\rm{BS}2}})t]},\label{tworealrootslong2}
\end{eqnarray}
whose periodic is $T=2\pi/({E}_{\rm{BS}1}-{E}_{\rm{BS}2})$. Here, ${ E}_{\rm{BS}1}$ and ${E}_{\rm{BS}2}$ are two pure imaginary roots of ${\cal G}(s=-i{ E}_{\rm{BS}})$, where these coefficients take ${\sigma_1} = 1/{{{\cal G}'( - i{ E}_{\rm{BS}1})}},{\sigma _2} =1/{{{\cal G}'( - i{E}_{\rm{BS}1})}}$. The dynamics reaches periodic oscillation behaviors.  In other words, the non-localized mode will approach zero after some time due to the localized exciton dynamics. The short-time dynamics are given by Fig.~\ref{Ut}(b). The changes from complete decoherence to decoherence suppression and then to periodic oscillation result from the existence of bound-states in the model itself.

(iii) {\it Complete gain of cavity amplitude}.
We now discuss in detail the features of the third regime for the cavity amplitude with non-Markovian gain in Fig.~\ref{Ut}. Firstly, the localized mode vanishes due to there being no real root; the exciton dynamics undergoes a full gain process (see Fig.~\ref{Ut}(c)). It can be approximately characterized as a non-localized mode, which contains two parts: one is the second term in Eq.~(\ref{calCt}), which is the oscillating gain process due to the complex roots in ${\cal L}(s)=0$ in the regime of [${\mathop{\rm Re}\nolimits} (s)<0$ and ${\mathop{\rm Im}\nolimits} (s)<\omega_e$]. The other is the integral part, i.e., the non-exponential parts will oscillate rapidly in time. This rapidly oscillating damping originates from the terms containing $e^{i\omega_et}$ in Eq.~(\ref{calCt}). In the two complex roots of Eq.~(\ref{appGx}) in Fig.~\ref{Ut}(c), the gain is generated. In this case, the effective gain reflected in Fig.~\ref{Ut}(c) is also different from that in Eq.~(\ref{jiage}) under the Markovian approximation, which originates from the nonexponential gain because of the non-Markovian effects in Eq.~(\ref{calCt}).

\section{The non-Markovian master equation of driven cavity}\label{section IV}
\subsection{Characteristic function method for the reduced density matrix}
In this section, we adopt the characteristic function method to obtain the exact non-Markovian master equation of the driving cavity. The characteristic function of the reduced density matrix represents the mean value of the cavity displacement operator under symmetric ordering:
\begin{equation}
\begin{aligned}
{Z} (\eta ,t) = \Tr[{e^{\eta {{\hat a}^\dag }(t) - {\eta ^*}\hat a(t)}}{\rho _T}(0)],
\label{chit}
\end{aligned}
\end{equation}where ${\rho _T}(0)$ denotes the initial state for the total system. Let us assume that the system and the environment initially are uncorrelated. The environment modeled by the Hamiltonian ${\hat H_R} = \sum\nolimits_k {{\omega _k}\hat b_k^\dag {\hat b_k}} $ is in a state of thermal equilibrium
\begin{equation}
\begin{aligned}
{\rho _{T}}\left( 0 \right) = {\rho _S} \otimes {\rho _R},{\rho _R} = \frac{{{e^{ - \beta {{\hat H}_R}}}}}{{{\rm{T}}{{\rm{r}}_R}{e^{ - \beta {{\hat H}_R}}}}},
\label{thermalequilibriumstate}
\end{aligned}
\end{equation}while the system is in a coherent state $\left| \alpha  \right\rangle $, where ${\rho _S}{\rm{ = }}\left| \alpha  \right\rangle \langle \alpha |$ can be obtained by defining it as an eigenstate of the annihilation operator $\hat a$ with an eigenvalue $\alpha$, $\beta  = 1/{\kappa _B}T$ with $\kappa_B$ is the Boltzmann constant, and $T$ the temperature of the environment. This work only focuses on the case of the zero-temperature. Substituting Eq.~(\ref{eq11c}) into Eq.~(\ref{chit}), we obtain
\begin{equation}
\begin{aligned}
{Z} (\eta ,t) = {Z} (\eta {{\cal U}^*},0)\mu (\eta ,t),
\label{chituvk}
\end{aligned}
\end{equation}
where $\mu (\eta ,t) = \exp [ {\eta {\varepsilon ^*}(t) - {\eta ^*}\varepsilon (t) - {y_1}(t){{\left| \eta  \right|}^2}/2} ]$, $\varepsilon (t) =  - i\int_0^t {F(\tau )} {\cal U}(t - \tau )d\tau $, $y_1(t)=\sum_k {|{v_k}(t){|^2}}$, and ${v_k}(t) = g_k^*\int_0^t {{\cal U}(t - \tau ){e^{i{\omega _k}\tau }}} d\tau$. Defining $\tilde {Z} (\eta ,t) = {{{Z} (\eta ,t)}}/{{\mu (\eta ,t)}}$ and applying Eq.~(\ref{chituvk}) for ${Z} (\eta ,t)$ and $\tilde {Z} (\eta ,0) = {Z} (\eta ,0)$, we can write $\tilde {Z} (\eta ,t)$ at time $t$ in terms of $\tilde {Z} (\eta ,0)$ as
\begin{equation}
\begin{aligned}
\tilde {Z} (\eta ,t) = \tilde {Z} \left( {\eta {{\cal U}^*},0} \right).
\label{chi5}
\end{aligned}
\end{equation}Due to $\tilde {Z} (\eta ,t)$ depending on $\eta$, ${\eta ^ * }$, and $t$ only through $\eta {{\cal U}^ * }$ and ${\eta ^ * }{\cal U} $ and differentiating Eq.~(\ref{chi5}) with respect to time, we have
\begin{equation}
\begin{aligned}
\frac{{\partial \tilde {Z} }}{{\partial t}} = {\text{ }}\frac{{\partial \tilde {Z} }}{{\partial (\eta {{\cal U}^*})}}\eta {{\dot {\cal U}}^*} + \frac{{\partial \tilde {Z} }}{{\partial \left( {{\eta ^*}u} \right)}}{\eta ^*}\dot {\cal U}.
\label{chi6}
\end{aligned}
\end{equation}For derivatives with respect to $\eta$ and ${\eta ^ * }$, we obtain
\begin{equation}
\begin{aligned}
\frac{{\partial \tilde {Z} }}{{\partial \eta }} = \frac{{\partial \tilde {Z} }}{{\partial \left( {\eta {{\cal U}^*}} \right)}}{{\cal U}^*}, \ \frac{{\partial \tilde {Z} }}{{\partial {\eta ^*}}} = \frac{{\partial \tilde {Z} }}{{\partial \left( {{\eta ^*}{\cal U}} \right)}}{\cal U} .
\label{chi7}
\end{aligned}
\end{equation}Solving the last two equations for the derivatives of ${\tilde {Z} }$ with respect to $\eta {{\cal U}^ * }$ and ${\eta ^ * }{\cal U}$ and substituting them into Eq.~(\ref{chi6}), we obtain a closed equation for ${\tilde {Z} }$
\begin{equation}
\begin{aligned}
\frac{{\partial \tilde {Z} }}{{\partial t}} = {\xi ^*}(t)\eta \frac{{\partial \tilde {Z} }}{{\partial \eta }} + \xi (t){\eta ^*}\frac{{\partial \tilde {Z} }}{{\partial {\eta ^*}}},
\label{chi8}
\end{aligned}
\end{equation}where $\xi (t) = {\dot {\cal U}}/{{\cal U}}$. Substituting $\tilde {Z} (\eta ,t) = {{{Z} (\eta ,t)}}/{{\mu (\eta ,t)}}$ into Eq.~(\ref{chi8}) we obtain
\begin{align}
\frac{{\partial {Z} }}{{\partial t}} = &+{\xi ^*}(t)\eta \left[ {\frac{{\partial {Z} }}{{\partial \eta }} - \left( {\frac{{\partial \ln \mu }}{{\partial \eta }}} \right){Z} } \right] + \left( {\frac{{\partial \ln \mu }}{{\partial t}}} \right){Z} \nonumber\\&+ \xi (t){\eta ^*}\left[ {\frac{{\partial {Z} }}{{\partial {\eta ^*}}} - \left( {\frac{{\partial \ln \mu }}{{\partial {\eta ^*}}}} \right){Z} } \right].\label{chi9}
\end{align}
The equation for the characteristic function ${Z}(\eta, t)$ forms a closed-form equation. The explicit structure of the time-dependent operator is dictated by the function $\mu(\eta, t)$, which is contingent upon the initial state of the environment. Substituting $\mu(\eta, t)$ into Eq.~(\ref{chi9}) gives
\begin{equation}
\begin{aligned}
\frac{{\partial {Z} }}{{\partial t}} = &+{\xi ^*}(t)\eta \frac{{\partial {Z} }}{{\partial \eta }} + \xi (t){\eta ^*}\frac{{\partial {Z} }}{{\partial {\eta ^*}}} 
+ \kappa (t)|\eta {|^2}{Z} \\&+ {\sigma ^*}(t)\eta {Z}  - \sigma (t){\eta ^*}{Z},
\label{chi10}
\end{aligned}
\end{equation}
where coefficients $\kappa (t)$ and $\sigma (t)$ are given by
\begin{equation}
\begin{aligned}
\kappa (t) &= [{y_1}(t)\left( {\xi  + {\xi ^*}} \right) - {{\dot y}_1}(t)]/2,\\
\sigma (t) &=  - \xi \varepsilon (t) + \dot \varepsilon (t).
\label{kappat}
\end{aligned}
\end{equation}

\subsection{Non-Markovian master equation for the cavity}
\begin{figure}[t]
\centerline{
\includegraphics[width=8.6cm, height=10cm, clip]{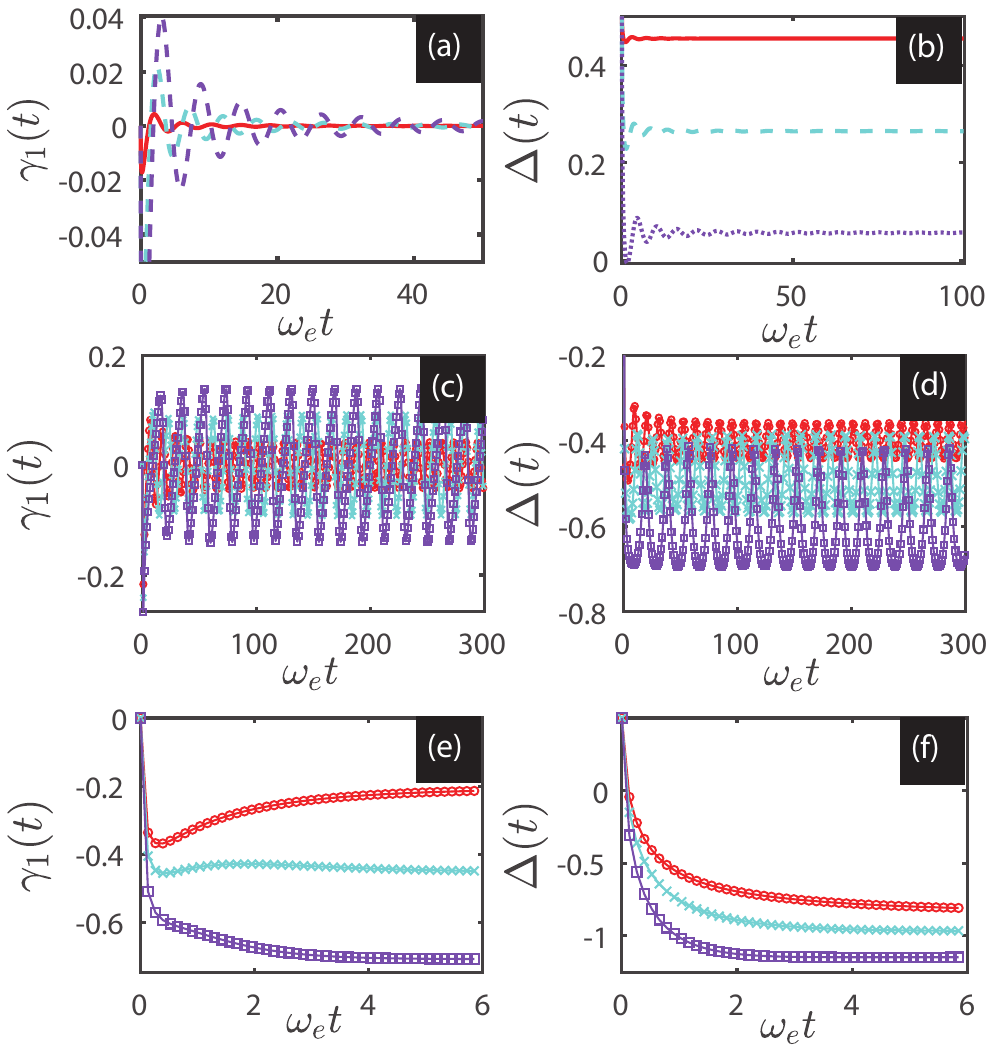}}
\caption{ Time evolution of $\Delta(t)$ and $\gamma_1(t)$ given by Eq.~(\ref{eq63c}). The parameters chosen are $\omega_l = \omega_e$, $f = \omega_e$. The different lines correspond to $\Gamma = 0.1\omega_e$, $\Gamma = 0.5\omega_e$, $\Gamma = 0.9\omega_e$ for (a) and (b); $\Gamma = 1.6\omega_e$, $\Gamma = 1.7\omega_e$, $\Gamma = 1.8\omega_e$ for (c) and (d); $\Gamma = 2.0\omega_e$, $\Gamma = 2.4\omega_e$, $\Gamma = 3.0\omega_e$ for (e) and (f). The other parameters are the same as in Fig.~\ref{Ut}.} \label{Ggamma1_Delta_t}
\end{figure}
\begin{figure}[t]
\centerline{
\includegraphics[width=8.4cm, height=9.5cm, clip]{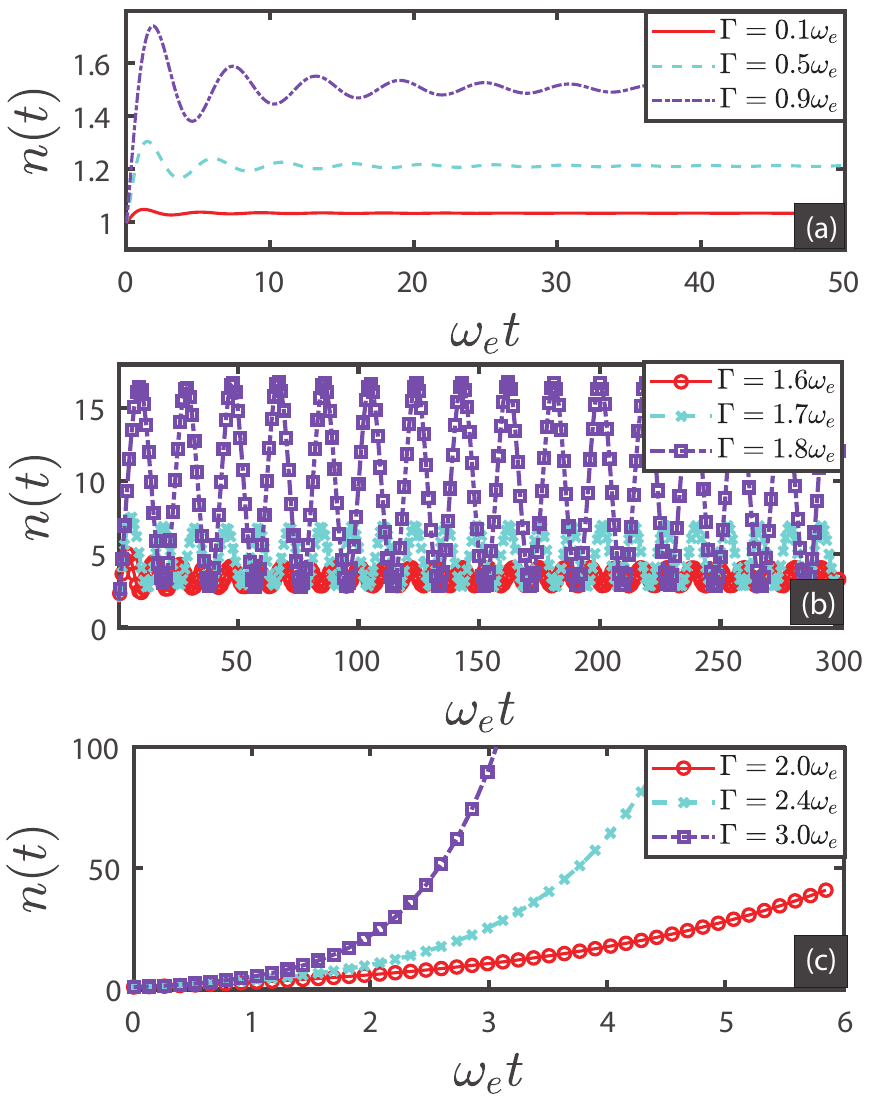}}
\caption{Time evolution of the average number of photons $n(t)$ given by Eq.~(\ref{eq48}), where ${{n}_0} = 1$, $\omega_l = 0$, and $f=0$. The other parameters are the same as in Fig.~\ref{Ut}.} \label{n(t)}
\end{figure}
\begin{figure}[t]
\centerline{
\includegraphics[width=8.4cm, height=10cm, clip]{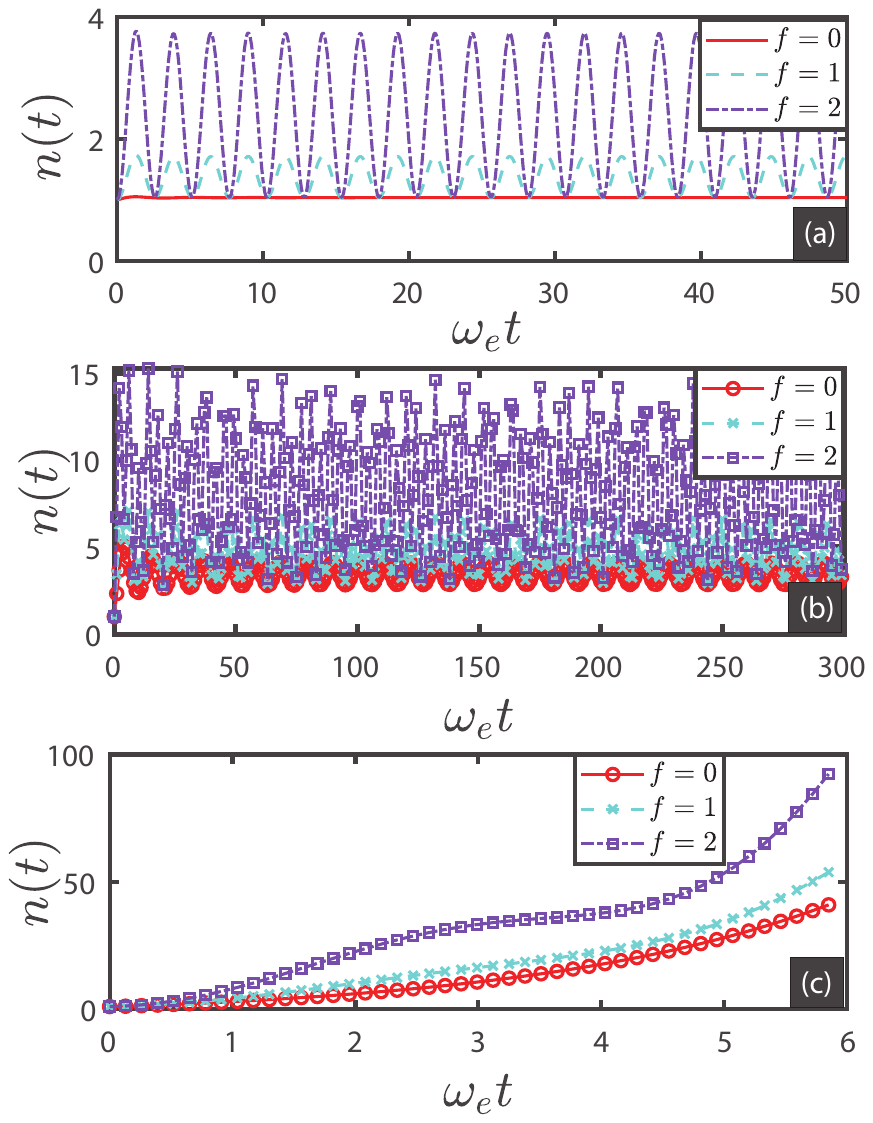}}
\caption{Time evolution of the average number $n(t)$ of photons with driving term given by Eq.~(\ref{eq48}). (a), (b), and (c) satisfy $\omega_l = 2\omega_e$, $\Gamma = 0.1\omega_e$, $\Gamma = 1.6\omega_e$, and $\Gamma = 2.0\omega_e$, respectively. The other parameters are the same as in Fig.~\ref{n(t)}.}
\label{fig4}
\end{figure}

\begin{figure}[t]
\centerline{
\includegraphics[width=8.4cm, height=8cm, clip]{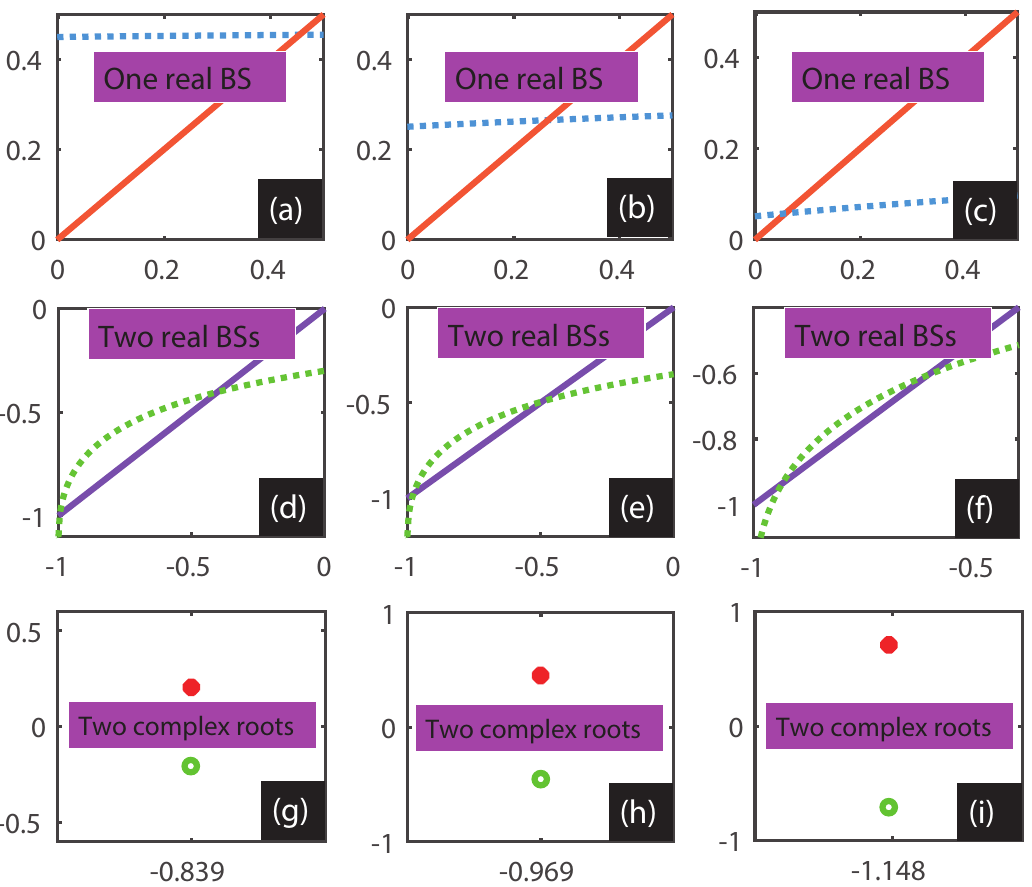}}
\caption{The intersection points in (a)-(f) are the solutions of the transcendental equation (\ref{eqvv10}) corresponding to the pure imaginary roots (bound states), while the complex roots of (g)-(i) are obtained from the solution of ${\cal L}(s)$ in Eq.~(\ref{1appLx}). The parameters chosen are $\Gamma = 0.1\omega_e$, $\Gamma = 0.5\omega_e$, $\Gamma = 0.9\omega_e$ for (a)-(c); $\Gamma = 1.6\omega_e$, $\Gamma = 1.7\omega_e$, $\Gamma = 1.8\omega_e$ for (d)-(f); $\Gamma = 2.0\omega_e$, $\Gamma = 2.4\omega_e$, $\Gamma = 3.0\omega_e$ for (g)-(n). The other parameters are the same as in Fig.~\ref{Ut}. The points of intersection of the red-solid, green-dashed, and blue-dashed lines denote the bound states in the system.} \label{energy}
\end{figure}

The full solution for single-mode cavity dynamics is determined by Eqs.~(\ref{chi10}) and (\ref{kappat}), where the time-dependent functions are ${\cal U}(t)$ and $\varepsilon (t)$, which is obtained by postulating an explicit driving cavity for the environment. In this model, the environment is regarded as a collection of other modes coupled to the cavity under study. Our objective is to derive a dynamical equation. The solution of this equation should exactly match Eq.~(\ref{chi10}), which should only involve the degrees of freedom of the system without the need to consider any others. To get this, we transform it from phase space back to Hilbert space with Eq.~(\ref{chi10}) by following the approach in Ref.~\cite{Walls1994}. Subsequently, we can obtain the exact non-Markovian master equation for the driving cavity
\begin{align}
\dot \rho (t) =&  - i[{\hat H_e}(t),\rho ] + {\gamma _1}(t)(2a\rho {\hat a^\dag } - {\hat a^\dag }a\rho  - \rho {\hat a^\dag }a)\nonumber\\
 &+ {\gamma _2}(t)(a\rho {\hat a^\dag } + {\hat a^\dag }\rho a - {\hat a^\dag }a\rho  - \rho a{\hat a^\dag }),
\label{eq37z}
\end{align}
where the time-dependent effective Hamiltonian
\begin{eqnarray}
\begin{aligned}
{{\hat H}_e}(t) =& \Delta (t){{\hat a}^\dag }\hat a + \phi (t)\hat a + {\phi ^*}(t){{\hat a}^\dag},
\label{eq36}
\end{aligned}
\end{eqnarray}
with
\begin{align}
{\gamma _1}(t) = - {\mathop{\rm Re}\nolimits} [\frac{{\dot {\cal U}(t)}}{{{\cal U}(t)}}],
\ \Delta (t) =  - {\mathop{\rm Im}\nolimits} [\frac{{\dot {\cal U}(t)}}{{{\cal U}(t)}}],\nonumber\\
\phi (t) =   - i[{\dot M^*}(t) - {\dot {\cal U}^*}(t){M^*}(t)/{{\cal U}^*}(t)],\label{eq63c}\\
{\gamma _2}(t) =  - 2{y_1}(t){\rm{Re}}[\dot {\cal U}(t)/{\cal U}(t)]+{\dot y_1}(t),\nonumber
\end{align}
and
\begin{align}
M(t) =&  - i\int_0^t {F(\tau )} {\cal U}(t - \tau )d\tau\nonumber\\
{y_1}(t) =&\int_0^t {d\tau \int_0^t {d\tau 'g(\tau ' - \tau )} } {{\cal U}^*}(t - \tau ){\cal U}(t - \tau ').\label{eq65}
\end{align}
Equations (\ref{eq11c}) and~(\ref{eq63c}) give ${y_1}(t)=|{\cal U}(t)|^2-1$ and ${\gamma _2}(t)=-2{\gamma _1}(t)$, which causes Eq.~(\ref{eq37z}) to become
\begin{align}
\dot \rho (t) =  - i[{\hat H_e}(t),\rho ] -{\gamma _1}(t)(2{\hat a^\dag }\rho \hat a - \hat a{\hat a^\dag }\rho  - \rho \hat a{a^\dag }),
\label{eq37cc}
\end{align}
which is a pure non-Markovian gain effect induced by the counterrotating-wave interactions in the cavity and structured environment. The average number $n(t)$ of photons with driving terms is
\begin{eqnarray}
\begin{aligned}
\!\!\!\!\!\!\!\!\!\! \langle \hat a^{\dag}(t) \hat a(t)\rangle
=|{\cal U}(t){|^2}n(0)+|{\cal U}(t)|^2-1+M^{*}(t)M(t).
\label{eq48}
\end{aligned}
\end{eqnarray}
Figure~\ref{Ggamma1_Delta_t} plots the relationship between $\gamma_1(t)$ and $\Delta(t)$ with time, which are given by Eq.~(\ref{eq63c}). We find that in Fig.~\ref{Ggamma1_Delta_t}(a), (c), and (e), as the value of $\Gamma$ increases, the amplitude of $\gamma_1(t)$ increases, and then Fig.~\ref{Ggamma1_Delta_t}(a) eventually tends to stabilize. Figure~\ref{Ggamma1_Delta_t}(c) achieves equal-amplitude oscillation, while Fig.~\ref{Ggamma1_Delta_t}(e) shows decay or gain. In Fig.~\ref{Ggamma1_Delta_t}(b), (d), and (f), as the value of $\Gamma$ increases, the value of $\Delta(t)$ decreases. In the beginning, the amplitude of Fig.~\ref{Ggamma1_Delta_t}(b) is weak and eventually tends to stabilize. Figure~\ref{Ggamma1_Delta_t}(d) shows equal-amplitude oscillations, while Fig.~\ref{Ggamma1_Delta_t}(f) shows decay.

Figure~\ref{n(t)} shows the relationship between the photon number $n(t)$ and time, which is similar to Fig.~\ref{Ut}. As $\Gamma$ increases, the $n(t)$ value increases. In Fig.~\ref{n(t)}(a), the lines eventually tend to stabilize. Equal-amplitude oscillations occur in Fig.~\ref{n(t)}(b). Figure~\ref{n(t)}(c) generates stronger gain. In Fig.~\ref{fig4}, we plot the relation between the average number of photons with driving term and time for different $f$ and $\Gamma$. When comparing Fig.~\ref{n(t)}(a) with Fig.~\ref{fig4}(a), the number of photons in the cavity changes from a stable steady-state solution to a curve of periodic equal-amplitude oscillation. This is because the contribution of the driving field in Eq.~(\ref{eq10}) undergoes quantum interference with a bound state solution. When looking at Fig.~\ref{n(t)}(b) with Fig.~\ref{fig4}(b), the number of photons in the cavity changes from periodic equal-amplitude oscillation to non-equal-amplitude periodic oscillation. This stems from the overlap between the two bound-state solutions of the photon number in Eq.~(\ref{eq48}) and the driving term, thus generating quantum interference of three exponents. Finally, when comparing Fig.~\ref{n(t)}(c) with Fig.~\ref{fig4}(c) instead of oscillation, the gain occurs. The reason is that there is no bound state for the photon number in Eq.~(\ref{eq48}), where the driving term has only one term and does not undergo interference with the bound state.

\subsection{The case of rotating wave approximation}
Considering the Hamiltonian that describes the coupling between the driven cavity and the structured environment under the rotating-wave approximation (RWA)~\cite{PhysRevLett.133.050401,PhysRevApplied.17.034073,PhysRevA.106.062438,PhysRevA.104.042609,PhysRevLett.109.170402}
\begin{align}
\hat H_{\rm{RWA}} (t)&=\Delta {{\hat a}^\dag }\hat a + \sum\nolimits_k {{\tilde {\Omega} _k}} \hat b_k^\dag {{\hat b}_k} + F(t){{\hat a}^\dag } + {F^*}(t)\hat a \nonumber\\
&+i\sum\nolimits_k {( {g_k^*{{\hat a}^\dag }\hat b_k  - {g_k}\hat a{{\hat b}_k^\dag}} )}, \label{Hamiltonian-rwa}
\end{align}
we obtain the master equation with RWA as follows
\begin{equation}
\begin{aligned}
\!\!\!\!\dot \rho \left( t \right) \!= \! - i[ {{\hat H_{\rm{RWA}}}(t),\rho } ] \!-\!{\gamma _{\rm{RWA}}}(t)(2{\hat a }\rho \hat a^\dag \!-\! \hat a ^\dag {\hat a }\rho  \!-\! \rho \hat a ^\dag {\hat a }).\!\!\!\!
\label{master_equationz}
\end{aligned}
\end{equation}
Herein, we rewritten Eq.~(\ref{Hamiltonian-rwa}) as ${{\hat H}_{\rm{RWA}}}(t) =  {\Delta_{\rm{RWA}}} (t){{\hat a}^\dag }\hat a + {\phi_{\rm{RWA}}} (t)\hat a + {\phi _{\rm{RWA}}^*}(t){{\hat a}^\dag}$. Thus, by assuming that the environment is initially in the vacuum state, the coefficients $ \Delta \left( t \right)$, $ \gamma_{\rm{RWA}} \left( t \right)$, and $ \phi_{\rm{RWA}} \left( t \right)$ in the master equation can be determined by
\begin{align}
 {\Delta_{\rm{RWA}}} \left( t \right) &=  - {\rm{Im[}}\dot u(t)/u(t)],\
{ \gamma _{\rm{RWA}}}\left( t \right) =  - {\rm{Re}}[\dot u(t)/u(t)],\nonumber\\
 \phi_{\rm{RWA}} (t) &=  - i{{\dot x}^ * }(t) + i[{{\dot u}^*}(t)/{u^*}(t){{\dot x}^ * }(t)],
\label{coefficient2}
\end{align}
with the parameters satisfying
\begin{small}
\begin{equation}
\begin{aligned}
&\dot u =- i{\omega _c}u(t) - \int_0^t {g_{\rm{RWA}}(t - \tau )} u(\tau )d\tau ,\\
&x(t) =  - i\int_0^t {F (\tau )} u(t - \tau )d\tau,
\label{rwaxishu}
\end{aligned}
\end{equation}
\end{small}where $g_{\rm{RWA}}(t - \tau )=\int {J(\omega )} {e^{ - i\omega t}}d\omega$ denotes the correlation function corresponding to the rotating-wave interactions. Equation (\ref{master_equationz}) shows the exact non-Markovian master equation of the driving cavity under the rotating-wave approximation, which represents a cavity decay process caused by the environment. It is completely different from the exact non-Markovian master equation (\ref{eq37z}) with gain, which originates from the counterrotating-wave interaction corresponding to Hamiltonian (\ref{eq2}).

\section{Bound state and non-Markovian dynamics}\label{section V}
In this section, we will show that the non-Markovian gain for the cavity is due to the bound state of the whole system (cavity plus environment)~\cite{PhysRevE.93.012107,PhysRevX.6.021027,PhysRevA.81.052330,Zhang2013LonglivedQC,Kofman1994SpontaneousAI}. Possible realizations of the prediction can be observed within current technologies~\cite{Noda2007SpontaneousemissionCB,Liu2016QuantumEN,Lodahl2004ControllingTD}. To proceed,  we make a Laplace transform to $ {\cal{U}}(t)$ and obtain $ {\cal{U}}(s) = [s + i \Delta - {g}(s)]^{-1}$ with $g(s) = \int {{{J(\omega )} \mathord{\left/
 {\vphantom {{J(\omega )} {\left( {s - i\omega } \right)}}} \right.
 \kern-\nulldelimiterspace} {\left( {s - i\omega } \right)}}d\omega } $. According to the Cauchy residue theorem, the inverse Laplace transform can be done by finding the all poles of ${ {\cal U}}(s)$. We now consider a special case if there is a pole on the imaginary axis, i.e., purely imaginary axis $s=-iE_{\rm{BS}}$ (where $E_{\rm{BS}}$ is a real number),
in which poles equation
\begin{eqnarray}
\begin{aligned}
{\cal G}(s){|_{s =  - iE_{\rm{BS}}}}=s + i{\Delta} - g(s)  = 0
\label{eqvv7}
\end{aligned}
\end{eqnarray}corresponds to the bound state contributions (the first term in Eq.~(\ref{calCt}) with $E_{\rm{BS}} \equiv i{\chi} _m^{(1)}$) and leads to the identity
\begin{eqnarray}
\begin{aligned}
{\Delta} - \int_{\omega_e}^{ + \infty } {\frac{{J(\omega )}}{{\omega +E_{\rm{BS}} }}d\omega  = E_{\rm{BS}}},
\label{eqvv10}
\end{aligned}
\end{eqnarray}
which can also be regenerated from the eigenequation of the effective Hamiltonian derived from the Heisenberg equation in Appendix~\ref{section C}. The transcendental equation satisfied by the bound state given by Eq.~(\ref{eqvv10}) is completely different from that of the model under the rotating wave approximation, where Eq.~(\ref{eqvv10}) becomes $\Delta  - \int {{{J(\omega )} \mathord{\left/{\vphantom {{J(\omega )} {\left( {\omega  - {E_{{\rm{BS}}}}} \right)}}} \right. \kern-\nulldelimiterspace} {\left( {\omega  - {E_{{\rm{BS}}}}} \right)}}d\omega }  = {E_{{\rm{BS}}}}$~\cite{PhysRevLett.133.050401,PhysRevApplied.17.034073,PhysRevA.106.062438,PhysRevA.104.042609,PhysRevLett.109.170402,PhysRevA.93.033833,PhysRevX.6.021027}. The localized modes exist if and only if the environmental spectral density has band gaps located at the pure imaginary zeros with ${\cal G}(- i{\chi} _m^{(1)}) = 0$ (see point $P$ in Fig.~\ref{contour_g} (a)). These localized modes do not decay, which gives dissipationless non-Markovian dynamics.

In Fig.~\ref{energy}(a), (b), and (c), there is an intersection between the orange solid line and the blue dashed line, indicating the existence of a bound state with gain. In Fig.~\ref{energy}(d), (e), and (f), there are two intersections between the purple solid line and the green dashed line, at which point there are two bound states. In Fig.~\ref{energy}(g), (h), and (i), there are two intersection points, at which point two complex roots are satisfying Eq.~(\ref{1appLx}).

\section{The relationship between the reduced density matrices of the driving and non-driving cavity}\label{section VI}

The Hamiltonian without the driving term reads
\begin{align}
\hat H_1 = \Delta {{\hat a}^\dag }\hat a + \sum\limits_k {{\tilde {\omega} _k}} \hat b_k^\dag {{\hat b}_k} + i\sum\limits_k {( {g_k^*{{\hat a}^\dag }\hat b_k^\dag  - {g_k}\hat a{{\hat b}_k}} )},
\label{eq2z}
\end{align}
where the exact non-Markovian master equation for the cavity is derived as
\begin{align}
{\dot {\rho}}_1(t) = & - i[\Delta {{{\hat a}}^\dag }\hat a ,\rho_1 ] + {\gamma _1}(t)(2\hat a\rho_1 {{\hat a}^\dag } - {{\hat a}^\dag }\hat a\rho_1  - \rho_1 {\hat a^\dag }\hat a)\nonumber\\
 &+ {\gamma _2}(t)(\hat a\rho_1 {\hat a^\dag } + {\hat a^\dag }\rho_1 \hat a - {\hat a^\dag }\hat a\rho_1  - \rho_1 \hat a{\hat a^\dag }).
\label{eq333z}
\end{align}
Defining $\rho (t) \!=\! D(\alpha (t)){\rho _1}(t){D^\dag }(\alpha (t))$ and $D(\alpha (t)) = \exp \left[ {\alpha (t){{\hat a}^\dag } - {\alpha ^*}(t)\hat a} \right]$, we can arrive at Eq.~(\ref{eq37z}), where $\alpha (t)$ is determined by $f_2(t)$ in Eq.~(\ref{eq21}). By comparing Eqs.~(\ref{eq37z}) and (\ref{eq333z}), we have an important insight: the influence of the external driving field is only to change the effective driving term in master equation (\ref{eq37z}), which does not affect the free term $\Delta(t)$ of the cavity and the dissipation terms $\gamma_1(t)$ as well as $\gamma_2(t)$. Moreover, we show that the time-coefficient $\phi (t)$ in Eq.~(\ref{eq63c}) induced by the driving field also contains the non-Markovian effect of the structured environment feedback on the cavity, which is specifically manifested in the cavity amplitude ${\cal U}(t)$ in Eqs.~(\ref{eq63c}) and (\ref{eq65}).

\section{transient current}\label{section VII}
The transient current from the system flow into the environment is defined in the Heisenberg picture as
\begin{eqnarray}
I(t) = \frac{d}{{dt}} \langle {\hat N(t)} \rangle  = -i\langle {[\hat N(t),\hat H(t)]} \rangle , \label{Itt}
\end{eqnarray} where $\hat N(t) = \sum\nolimits_k {\hat b_k^\dag (t)} {\hat b_k}(t)$. By explicitly calculating the above commutation relation with the Hamiltonian of Eq.~(\ref{eq2}), we obtain the transient equation for conservation current
\begin{eqnarray}
\frac{{\partial  n(t)}}{{\partial t}} =  S(t) +  I(t),\label{current}
\end{eqnarray} where $ n(t) = \Tr_S[{{\hat a}^\dag }\hat a\rho (t)]$ is the total exciton number in the cavity with $\rho (t)$ given by Eq.~(\ref{eq37z}), and $S(t) =  - i\Tr_{SR}[F (t)\langle {{\hat a^\dag }(t)} \rangle - {F ^ * }(t)\langle {\hat a(t)} \rangle]$ is the source coming from the driving field. Eq. (\ref{current}) indicates that the increase of the photon number in the resonators equals the photons received from the driving field, and the non-rotating wave term subtracts the photons flow into the environment.

\begin{figure}[htp!]
\centering
\includegraphics[width=8.6cm, height=10cm, clip]{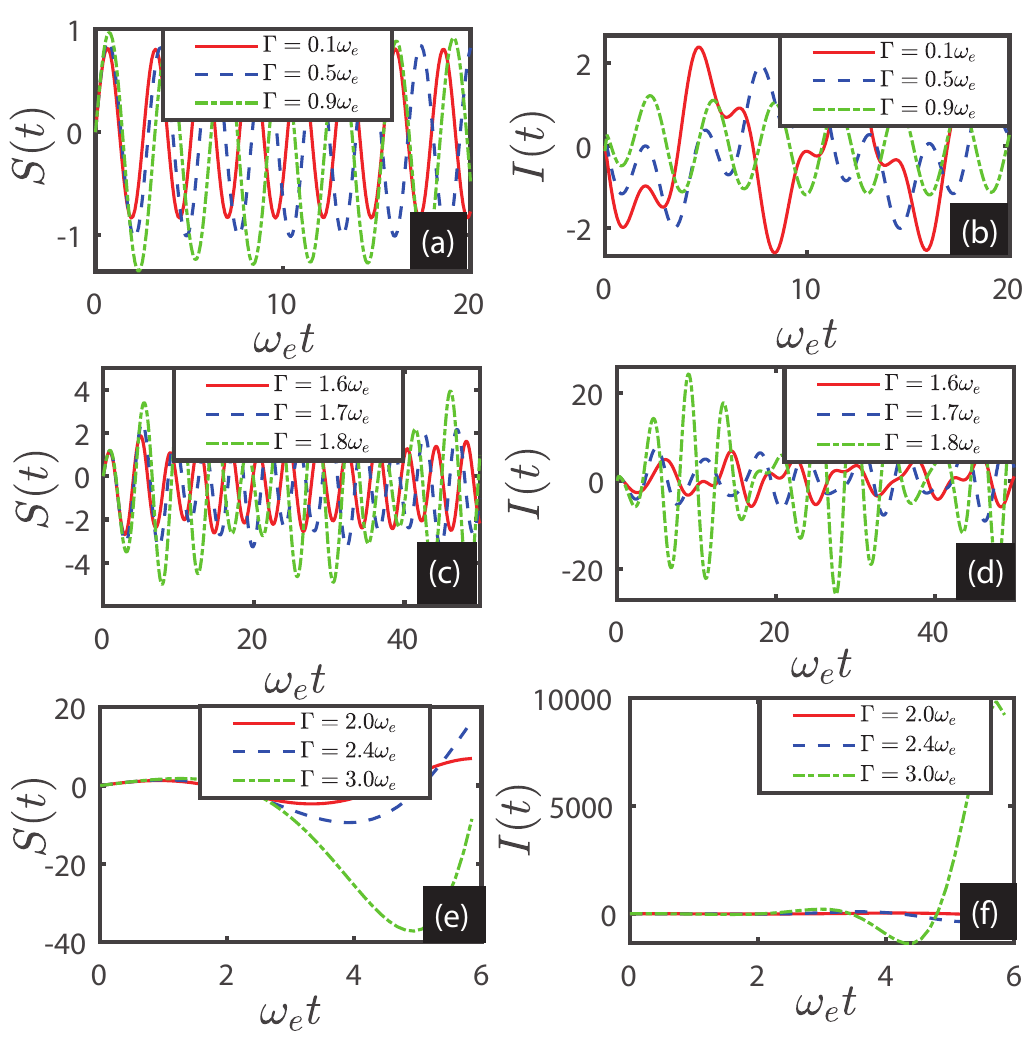}
\caption{(a)-(c) and (d)-(f) are current $S(t)$ coming from the driving field source, transient current $I(t)$ from the system flow into the environment versus time in Eq.~(\ref{current}). In this case, we choose $\Gamma = 0.1\omega_e$ (solid-line), $\Gamma = 0.5\omega_e$ (dashed-line), $\Gamma = 0.9\omega_e$ (dashed-dotted-line) for (a), and (d); $\Gamma = 1.6\omega_e$, $\Gamma = 1.7\omega_e$, $\Gamma = 1.8\omega_e$ for (b), and (e); $\Gamma = 2.0\omega_e$, $\Gamma = 2.4\omega_e$, $\Gamma = 3.0\omega_e$ for (c), and (f). Here $\omega_l = 2\omega_e$, $f = \omega_e$. The other parameters are the same as in Fig.~\ref{n(t)}.} \label{S+I}
\end{figure}
\begin{figure}[t]
\centerline{
\includegraphics[width=6.8cm, height=6.8cm, clip]{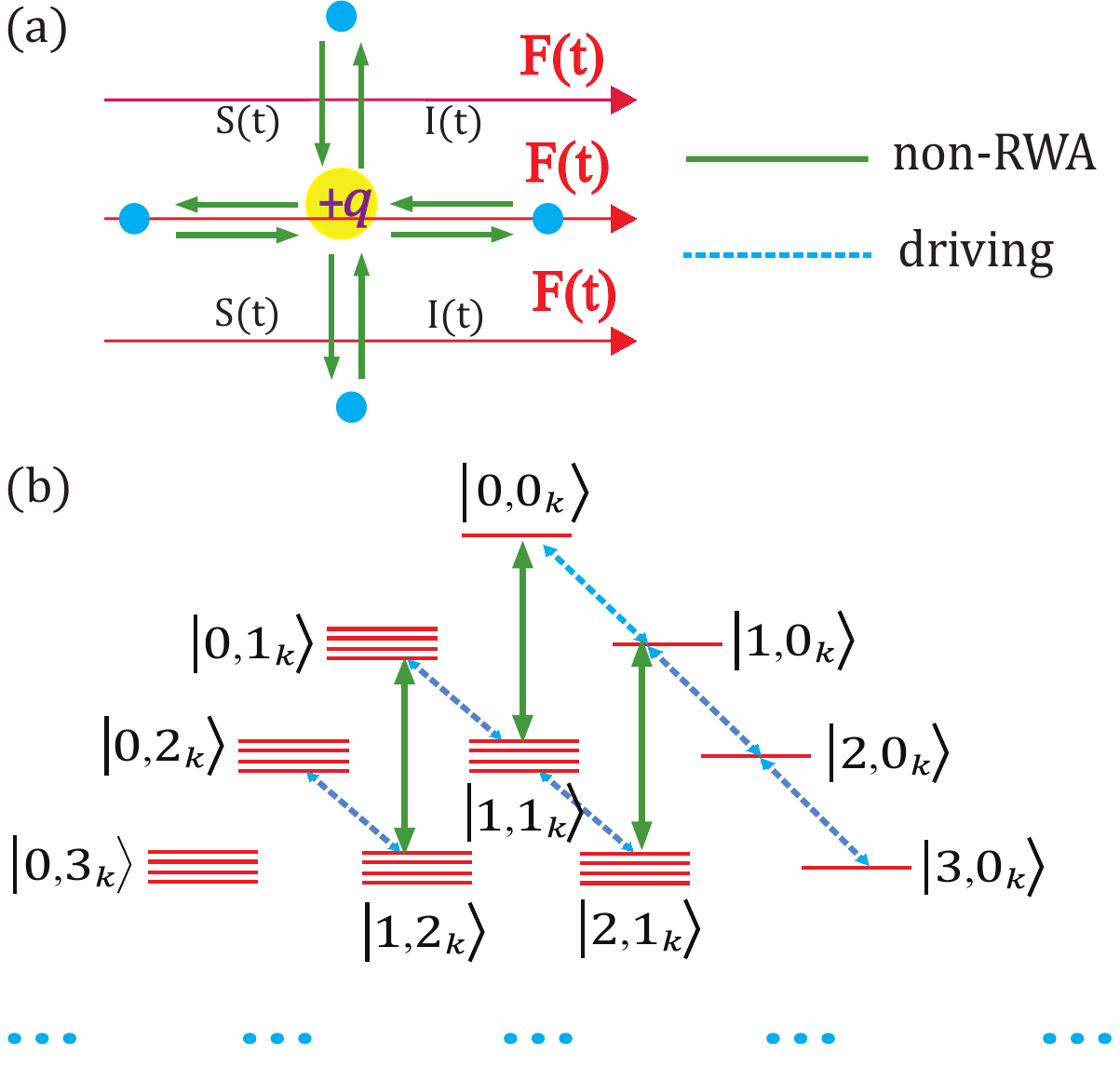}}
\caption{(a) In the framework of the factorized direct product of the system and the environment state, the conserved currents connected with each part are established. (b) Energy diagram showing the zero-exciton state, one-exciton state, and two-exciton state $\cdot$~$\cdot$~$\cdot$ $m_k+n$ exciton states $\left| {n,{m_k}} \right\rangle $ ($n$ excitons in the cavities, $m_k$ photons in $k$-th mode for the environment), and the transition paths: green solid lines with arrows denotes two-exciton transitions (non-RWA processes), blue-dotted lines with arrows denotes coherent driving sources (coherent processes).} \label{tu6}
\end{figure}

\begin{figure}[htp!]
\centering
\includegraphics[width=8.6cm, height=10cm, clip]{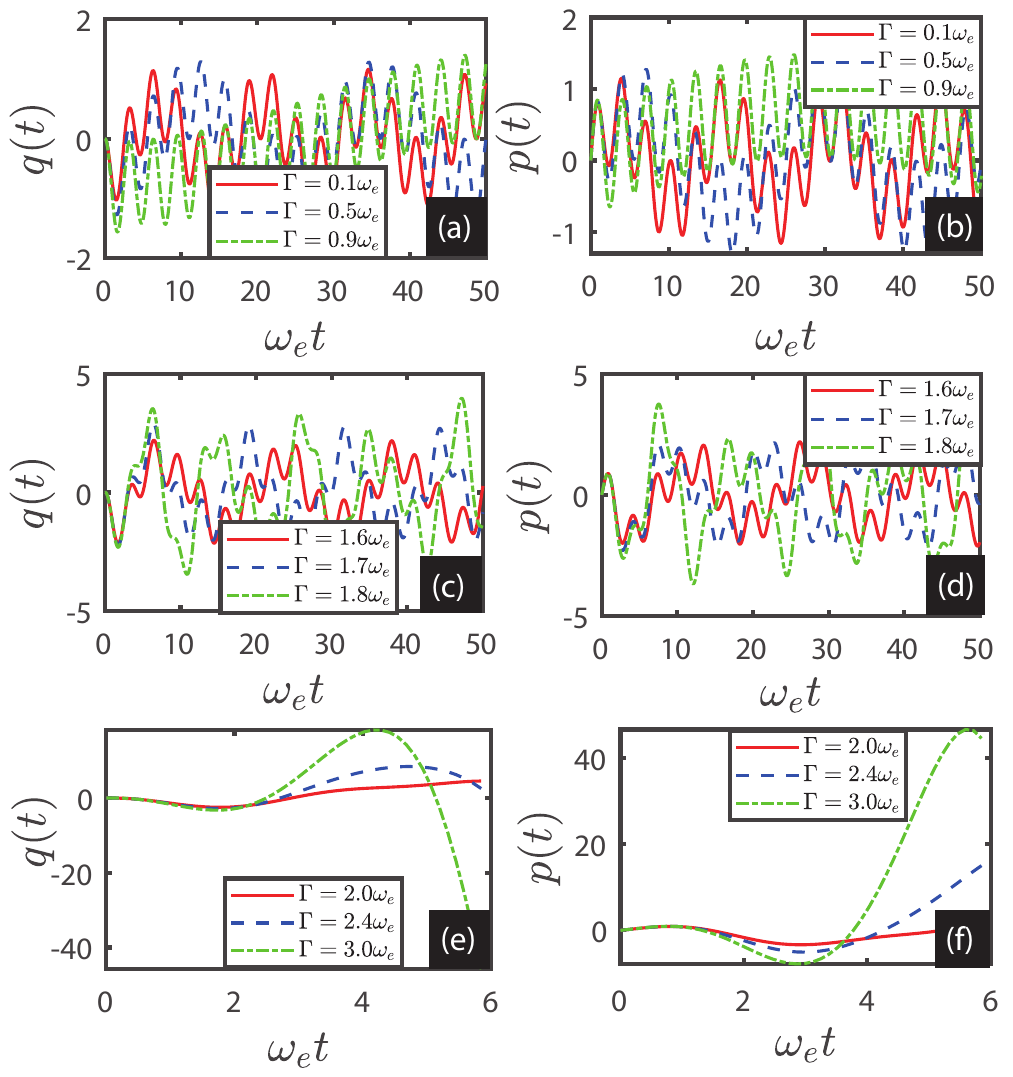}
\caption{Evolution in time of the expectation value of (a)-(c) position $p(t)$ and (d)-(f) momentum $q(t)$ [see Eq.~(\ref{vacuum})] in quantum vacuum state. The different lines correspond to different values of the driving field intensity. We set $\omega_l = 2\omega_e$, $f = \omega_e$, ${M_0}=\omega_e$, and $\omega_c = \omega_e$. The other parameters and vertical ordinates are given in Fig.~\ref{S+I}.} \label{tu7}
\end{figure}

Now, we turn to the exact numerical calculation and compare the above transient current in the weak coupling limit ($\Gamma  \ll {\omega _c}$) with the exact numerical solution of coherent driving sources, the transient current induced by the change in the number of photons, the two-photon current from the counterrotating-wave term, respectively. The result is plotted in Fig.~\ref{S+I}, where the cavity frequency $\omega_c=\omega_e$ and $\Gamma = 0.1\omega_e$, which belongs to the weak coupling. As the driving field intensity increases, the amplitude of the current increases. Figs.~\ref{S+I} (a), (c), and (e) correspond to the weak coupling case, while the coherence current $S(t)$ comes from the driving field source. This is because the coherence current $S(t)$ forces the cavity to tend to coherence. But the transient current $I(t)$ behaves the small time-dependent oscillations at the small dissipation $\Gamma$ in Fig.~\ref{S+I} (b) and (d), while Fig.~\ref{S+I}(f) appears huge and even divergent oscillations for large gains. In this case, the physical mechanism originates from the transition paths of blue-solid lines in Fig.~\ref{tu6} (b). This means the transient current can capture valuable information and play two roles. One is that counterrotating-wave interactions make the cavity produce the gain, while the other is that the structured environment causes the cavity to dissipate with non-Markovian effects.

The advantage of solving the equations of motion in the Heisenberg picture is that they easily allow us to compute the expected values of relevant operators. We define $\hat a = \hat X/\sqrt {2/{M_0}{\omega _c}}  - i\hat P/\sqrt {2{M_0}{\omega _c}}$, which can get 
$\hat q = (\hat a + {{\hat a}^\dag })\sqrt {2\hbar /{M_0}{\omega _c}} /2$ and $\hat p = ({{\hat a}^\dag } - \hat a)\sqrt {2\hbar /{M_0}{\omega _c}} /2i$. The expectation values for $\hat q$ and $\hat p$ are
\begin{align}
q(t) = \Tr_S[\hat q{\rho _S}(t)], \ p(t) = \Tr_S[\hat p{\rho _S}(t)].
\label{qp}
\end{align}
Herein, we note that the evolution of the position variance ${\sigma _{{q^2}}}(t) = \Tr_S[{{\hat q}^2}{\rho _S}(t)] - {q^2}(t)$ is obtained by squaring Eq.~(\ref{qp}) and taking the expectation value and similarly for the momentum variance ${\sigma _{{p^2}}}(t) = \Tr_S [{{\hat p}^2}{\rho _S}(t)] - {p^2}(t)$ and the position-momentum covariance
${\sigma _{qp}}(t) = \Tr_S [\{ q,p\} {\rho _S}(t)]/2 - q(t)p(t)$. In the vacuum state, we obtain
\begin{equation}
\begin{aligned}
{q(t}) = &[{M^*}(t) + M(t)]\sqrt {2\hbar /{M_0}{\omega _c}} /2,\\
{p(t}) =  &[{M^*}(t) - M(t)]\sqrt {2\hbar /{M_0}{\omega _c}} /2i,\\
{\sigma _{{q^2}}}(t) =  &[(2n(0) + 2)|{\cal U}(t){|^2} - 1](\hbar /2{M_0}{\omega _c}),\\
{\sigma _{{p^2}}}(t) = & [(2n(0) + 2)|{\cal U}(t){|^2} - 1](\hbar {M_0}{\omega _c}/2),\\
{\sigma _{qp}}(t) =  &0,
\label{vacuum}
\end{aligned}
\end{equation}while in the coherent state
\begin{align}
&{q(t}) = [\alpha {\cal U}(t) + {\alpha ^*}{{\cal U}^*}(t) + {M^*}(t) + M(t)]\sqrt {2\hbar /{M_0}{\omega _c}} /2,\nonumber\\
&{p(t}) = [{\alpha ^*}{{\cal U}^*}(t) - \alpha {\cal U}(t) + {M^*}(t) - M(t)]\sqrt {2\hbar /{M_0}{\omega _c}} /2i,\nonumber\\
&{\sigma _{{q^2}}}(t) = [2|{\cal U}(t){|^2} - 1](\hbar /2{M_0}{\omega _c}), \ {\sigma _{qp}}(t) = 0\nonumber\\
&{\sigma _{{p^2}}}(t) = [2|{\cal U}(t){|^2} - 1](\hbar {M_0}{\omega _c}/2), 
\label{coherent}
\end{align}

\begin{figure}[htp!]
\centering
\includegraphics[width=8.6cm, height=7cm, clip]{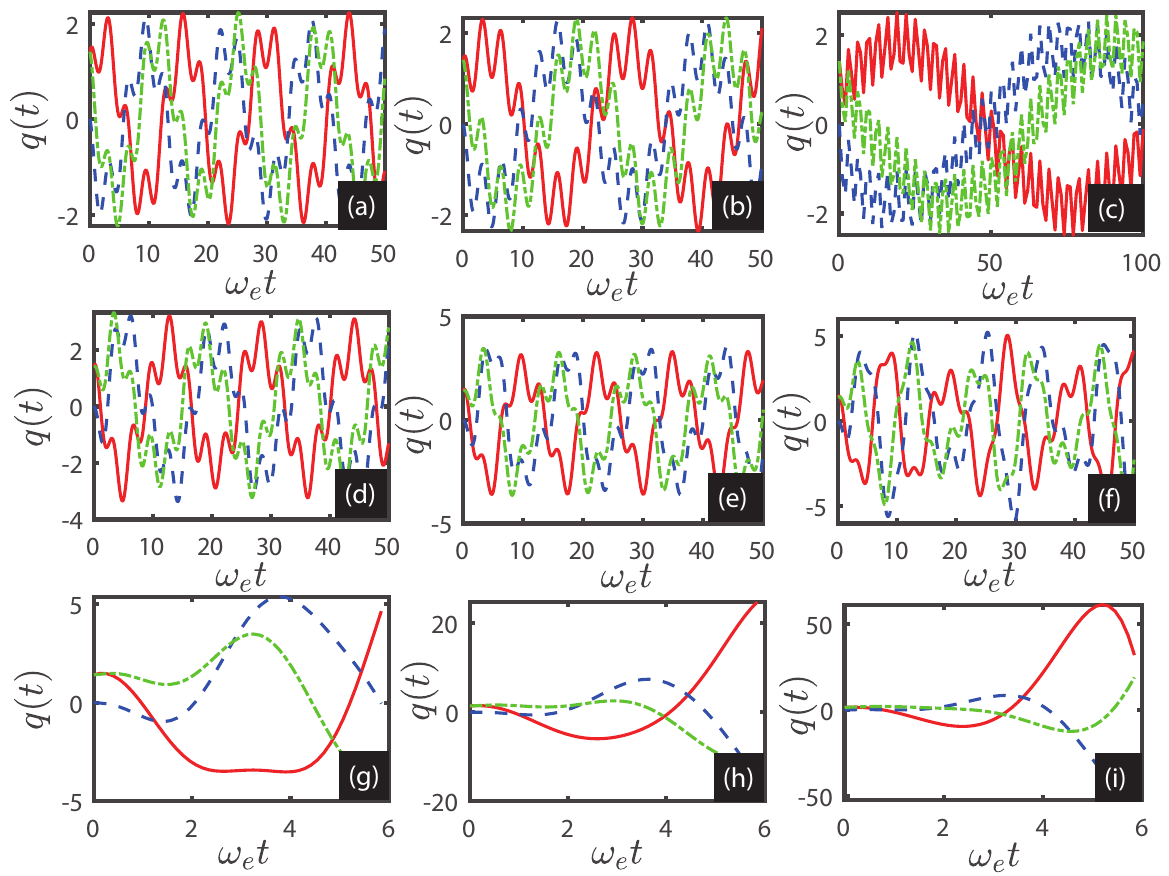}
\caption{Time evolution of the expectation value of position $p(t)$ in coherent state in Eq.~(\ref{coherent}). The different lines correspond to different initial coherent state: $\alpha= 1-i$ (red solid-line), $\alpha= i$ (blue dashed-line), and $\alpha= 1+i$ (green dashed-dotted-line). The other parameters are the same as in Fig.~\ref{tu7}.} \label{tu8}
\end{figure}
\begin{figure}[htp!]
\centering
\includegraphics[width=8.6cm, height=7cm, clip]{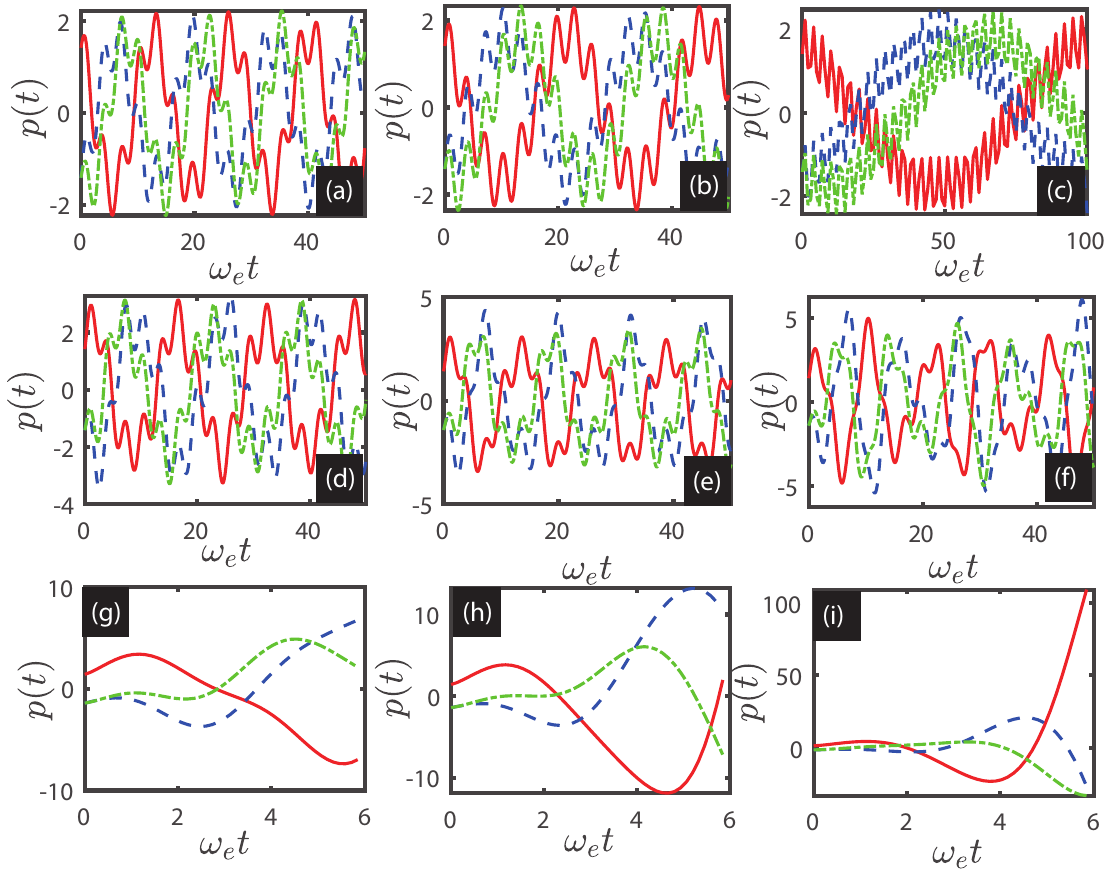}
\caption{Time evolution of the expectation value of momentum $q(t)$ in coherent state in Eq.~(\ref{coherent}), where the parameters are the same as in Fig.~\ref{tu7}.} \label{tu9}
\end{figure}

\begin{figure}[t]
\centerline{
\includegraphics[width=6.6cm, height=4cm, clip]{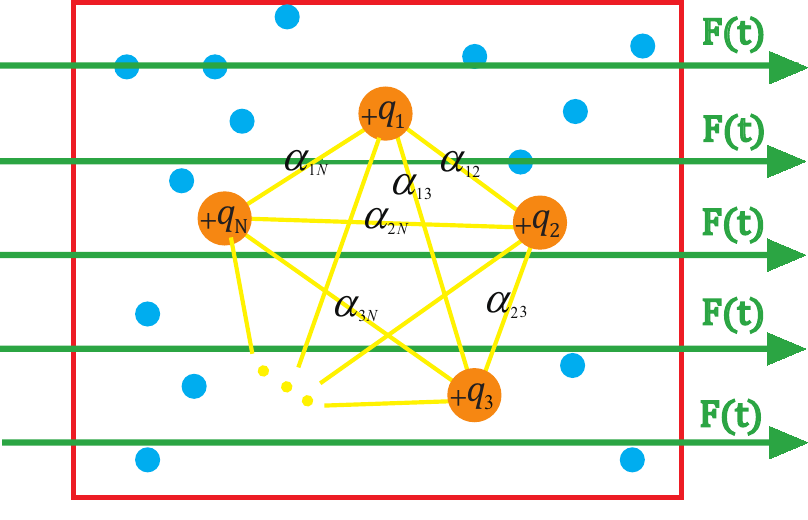}}
\caption{Quantum network consisting of $N$ mutually coupled cavities (coupled coefficients are is $\alpha_{mn}$) by counterrotating-wave interactions to the environment of harmonic oscillators with the frequency $\omega_k$. The orange-circle side represents the charged-Brownian oscillator, which is coupled to a large number of oscillators of the environment (bule circle) interacting via the coupling constants $g_{n,k}$.} \label{tu10}
\end{figure}

In Fig.~\ref{tu7}, we show the time evolution of the expectation values of position and momentum in a vacuum state for different values of $\Gamma$. We find periodic oscillations in position and momentum in Figs.~\ref{tu7}(a)-(d), but in Figs.~\ref{tu7}(e) and (f), position ultimately decays and momentum ultimately gains. Figures \ref{tu8} and \ref{tu9} correspond to the time variation of position and momentum in coherent states accompanied by different $\Gamma$ values. We obtain results similar to those in Fig.~\ref{tu7}. In the (a)-(f) plots of Figs.~\ref{tu8} and \ref{tu9}, there are still periodic oscillations in position and momentum. When $\alpha= 1-i$, position and momentum first decay and then gain, while in the other two cases, position and momentum both decay. Here, we do not plot the figures for variances in Eqs.~(\ref{vacuum}) and (\ref{coherent}) because they are similar to Fig.~\ref{Ut}. Moreover, we can also generalize these results to a more general network involving an arbitrary number of coupled cavities (see Fig.~\ref{tu10}).

\section{Summary}\label{section VIII}
In summary, we have studied the exact non-Markovian dynamics of counterrotating-wave interactions between a driving cavity and an anisotropic three-dimensional photonic crystal environment. An analytical solution for the exact cavity amplitude has been derived by inverse Laplace transform with contour integration of multivalued functions, which considers the contributions of both the bound state and the dissipative parts. Through the characteristic function method, we further derive the exact non-Markovian master equation of the cavity, which plays a crucial role in bringing about the gain of the cavity. We also offer an in-depth exploration of the physical origin of this gain in the context of the system, which is composed of the cavity and the environment and has bound states present. Our investigations reveal that three distinct types of bound states can be formed within the system. These include static bound states without photon number inversion, periodic equal-amplitude oscillations associated with two bound states, and gain phenomena related to two complex roots. Moreover, we formulate a current equation encompassing the source originating from the driving fields, the transient current induced by the change in the number of photons, and the two-photon current induced by the counterrotating-wave term. After that, the results we present are compared with those obtained through the rotating-wave approximation. These results are extended to a more general quantum network involving an arbitrary number of coupled cavities. The proposed formalism potentially paves the way for a more profound and accurate understanding of non-Markovian quantum networks.

Exploring non-Markovian quantum networks and the total excitation number of non-conserving systems beyond the rotating-wave approximation are important in quantum optics. This includes scenarios such as the non-rotating-wave interaction between a cavity and environment~\cite{PhysRevA.96.033805,PhysRevA.97.042121,PhysRevA.98.023856,PhysRevA.101.013826,PhysRevA.105.023707}
\begin{align}
\sum\limits_k {{\zeta _k}(\hat a + {{\hat a}^\dag })({{\hat c}_k} + \hat c_k^\dag )},\label{NRQA_I}
\end{align}
the anisotropic non-rotating-wave interaction~\cite{PhysRev.79.845,Frhlich1952InteractionOE,Nakajima1955PerturbationTI,Zheng2002DynamicsOA,PhysRevB.75.054302,PhysRevE.78.046608,PhysRevA.81.042116,PhysRevX.4.021046,PhysRevA.103.043708} 
\begin{align}
\sum\limits_k {{\zeta _k}({{\hat c}_k}{{\hat a}^\dag } + \hat c_k^\dag \hat a) + {\xi _k}({{\hat c}_k}\hat a + \hat c_k^\dag {{\hat a}^\dag })},\label{ANRWA_I}
\end{align}
and many body models~\cite{PhysRevLett.116.160401,PhysRevA.93.012106,PhysRevB.93.094205,PhysRevA.96.033828,PhysRevA.99.012106,PhysRevResearch.2.023214}. In general, total systems including the external environment beyond the rotating wave approximation (RWA) merit further investigation. Such explorations are not limited to the anisotropic non-RWA cases, but should also encompass more general coupling structures between different subsystems, which may take the form of $\sum\nolimits_{n,k} {({G_{n,k}}{{\hat C}_n}\hat D_k^\dag  + G_{n,k}^ * \hat C_n^\dag {{\hat D}_k} + {\zeta_{n,k}}{{\hat C}_n}{{\hat D}_k} + \zeta_{n,k}^ * \hat C_n^\dag \hat D_k^\dag } )$. Here, ${\hat C_n^\dag }$ (${{{\hat C}_n}}$) and ${\hat D_k^\dag }$ (${{{\hat D}_k}}$) are the creation (annihilation) operators for the total systems (including coupled cavities, classical driving field, and environment), while ${{G_{n,k}}}$ and ${{\zeta_{n,k}}}$ respectively represent the interacting strengths of the rotating-wave and non-rotating-wave coupling.

\section*{ACKNOWLEDGMENTS}
H. Z. S. was supported by National Natural Science Foundation of China under Grants No.~12274064, and Scientific Research Project for Department of Education of Jilin Province under Grant No.~JJKH20241410KJ. C. S. acknowledges financial support from the China Scholarship Council, the Japanese Government (Monbukagakusho-MEXT) Scholarship (Grant No.~211501), the RIKEN Junior Research Associate Program, and the Hakubi Projects of RIKEN. Y. H. Z. acknowledges the National Natural Science Foundation of China (NSFC)(Grants Nos.~12374333).

\appendix
\section{Discussions on the experimental implementation for the setup} \label{section A}
\begin{figure}[b]
\centering{
\includegraphics[width=6.38cm, height=6.6cm, clip]{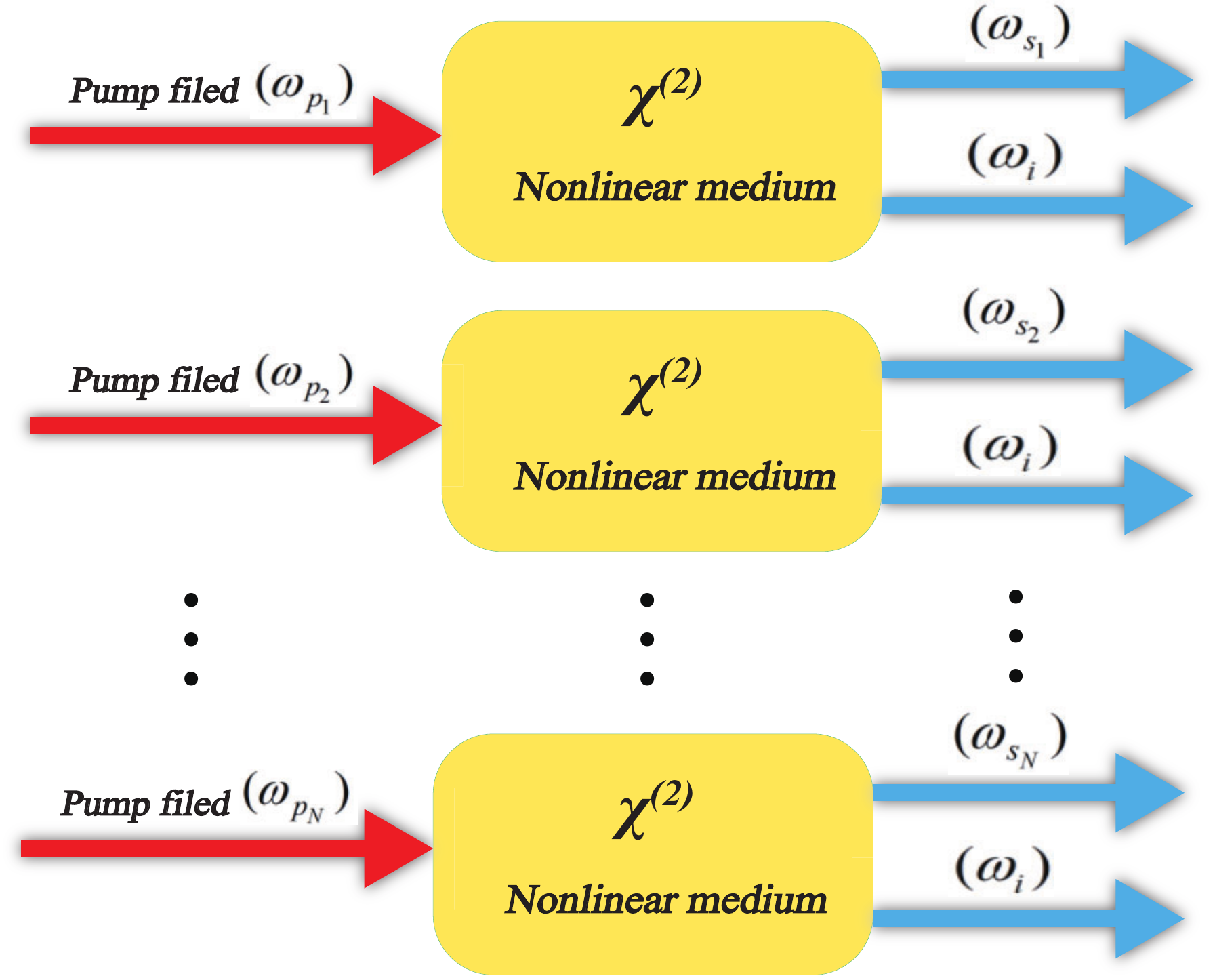}}
\caption{The experimental setup for the multiple nonlinear systems corresponding to Eq.~(\ref{Hsi1}) consists of $N$ optical parametric amplifiers (OPA). The figure represents the OPA process: A pump photon with frequency ${\omega _{{p_N}}}$ is down-converted into an identical pair of photons with frequency $\omega_{s_N}$ and $\omega_{i}$ after passing through the ${\chi}^{(2)}$ nonlinear medium.}\label{chicavity}
\end{figure}

We can realize the coupling of optical cavity with a degenerate optical parametric amplifier in parametric amplification system~\cite{Boyd2008,RevModPhys.82.1155,RevModPhys.84.1,PhysRevA.90.063824,Leghtas2014ConfiningTS,PhysRevA.94.033841,Minganti_2016,Adiyatullin2016PeriodicSI,PhysRevA.96.053810,PhysRevA.96.053827,Yan2018PulseregulatedSG,PhysRevA.100.023814} in Fig.~\ref{chicavity}, where the multiple nonlinear systems contain two photons (which denotes the signal ($\omega_{s_k}$) and idler ($\omega_i$), respectively), which originates from the frequency of a pump photon ($\omega_{p_k}$). Meanwhile, the above conversion follows the frequency relation $\omega_{p_k}=\omega_i+\omega_{s_k}$. This process is what we know about the parametric down-conversion in a dielectric medium with a ${\chi}^{(2)}$ nonlinearity. When the cavity is driven by a classical pump (such as a laser or a microwave generator), it is not significantly damped by the loss of photons via the down-conversion process. This Hamiltonian of the system can be written as
\begin{eqnarray}
{{\hat H}_g}{\rm{ = }}i\sum\limits_k {(g_k^*{{\hat a}^\dag }\hat b_k^\dag  - {g_k}\hat a{{\hat b}_k})},
\label{Hsi1}
\end{eqnarray}
where $g_k$ is the pump amplitude dependent upon coupling strength.

Moreover, the quantum counterrotating-wave interactions can be implemented in quantum optics setups or atomic Bose-Einstein Condensates. For quantum optics systems, we can use nonlinear wave mixing, such as spontaneous parametric down-conversion~\cite{PhysRevLett.18.732,Klyshko1970ParametricLA,PhysRevLett.128.173602} and spontaneous four-wave mixing~\cite{Du2008NarrowbandBG,PhysRevA.93.033815} through tunable designed parameters. The process of twin-beam generation into two modes due to spontaneous three- or four-wave mixing with a pump can also generate the two-mode squeezed fields~\cite{Caves1987QuantumWT,Shirasaki1990SqueezingOP,PhysRevA.73.063819,PhysRevA.100.043805}. The counterrotating-wave couplings for different cavities can be realized by matching the frequencies of cavities and the frequencies of the ac magnetic flux for the superconducting quantum interference devices (SQUID)~\cite{PhysRevB.87.014508,PhysRevLett.113.093602}, where the SQUID driven by external fluxes allows a modulation of the electrical boundary condition of the cavities and their couplings.

\section{The spectral density of photonic crystals environments}\label{section B}
The spectral density of the structured environment in Eq.~(\ref{eq8}) is given by
\begin{eqnarray}
{ J} (\omega ) = \sum\limits_{k } {{  G_{k}^* G_{ k}   }\delta (\omega  - {\Tilde {\omega }_k})},
\label{Jwdecided}
\end{eqnarray}
which describes the intensity of various noise frequency components in the environment. If the environmental spectral density has a relatively large value near the energy-level transition frequency of the cavity, it means that the environment has a significant impact on the decoherence process of the cavity. This is because the environment can interact more effectively with the cavity at this time, causing the cavity to lose its quantum coherence and resulting in the dissipation of quantum information into the environment. In the anisotropic three-dimensional photonic crystal~\cite{PhysRevA.50.1764}, the coupling coefficient $G_{k}$ is written as
\begin{equation}
{G_{k}} = ({\omega}{d}/\hbar )\sqrt {\hbar /(2{\varepsilon _0}{\Tilde {\omega }_k}V)} {\vec e_{k}} \cdot {\vec u}, \label{Vmk12}
\end{equation}
where $k$ denotes both the momentum and the polarization of the modes. ${d}$ and ${{\vec u}}$ describe the magnitude and unit vector of the cavity dipole moment of the transition. $V$ represents the quantization volume. ${{\vec e}_{k}}$ corresponds to the transverse unit vectors for the environment modes. ${{\varepsilon _0}}$ is the vacuum dielectric constant. According to Eq.~(\ref{Vmk12}), Eq.~(\ref{Jwdecided}) becomes
\begin{equation}
\begin{aligned}
{J} (\omega) =& \frac{{{  \varepsilon    }}}{{2{\varepsilon _0}\hbar V}}\sum\limits_{k} {\frac{{({{\vec e}_{k}} \cdot {{\vec u} })^2}}{{{\Tilde {\omega }_k}}}} \delta (\omega  - {\Tilde {\omega }_k}) \\
 =& \frac{{{  \varepsilon}}}{{2{\varepsilon _0}\hbar V}}\sum\limits_{k} {\frac{{1 - (\vec k \cdot {\vec u })^2/{k^2}}}{{{\Tilde {\omega }_k}}}}
 \delta (\omega  - {\Tilde {\omega }_k})\\
 =& \frac{{{\varepsilon}}}{{16{\pi ^3}{\varepsilon _0}\hbar }}\int {\frac{{[1 - (\vec k \cdot {\vec u })^2/{k^2}]{d^3}\vec k}}{{{\Tilde {\omega }_k}}}}
 \delta (\omega  - {\Tilde {\omega }_k}),
\label{TC1}
\end{aligned}
\end{equation}
where $\varepsilon=({\omega }{d })^2$, and we have replaced the sum by an integral via $\sum\nolimits_k { \to V {{(2\pi )}^{ - 3}}} \int {{d^3}\vec k}$
and $({{\vec e}_{k}} \cdot {{\vec u} })^2 = 1 - (\vec k \cdot {{\vec u} })^2/{k^2}$. Near the band edge, the dispersion relation may be expressed approximately by ${\Tilde {\omega }_k} = {\omega _e} + {B}{| {\vec k - \vec k_0 }|^2}$`\cite{PhysRevLett.58.2486}, where ${\omega }_e={\omega }_d+{\omega }_l$ (${\omega }_d$ denotes the cutoff frequency of band edge) is tunable because it depends on the frequency ${\omega }_l$ of driving field. We focus exclusively on the case where the cavity dipole moments are parallel~\cite{PhysRevA.69.023401}, a configuration that can be effectively emulated and has been experimentally realized, as demonstrated in Ref~\cite{PhysRevLett.89.163601}. Therefore, the angle between the dipole vector of the ${\vec k_0 }$ is all $\theta $. The angle between the dipole and ${\vec k}$ near ${\vec k_0 }$ is replaced approximately by $\theta $. Consequently, Eq.~(\ref{TC1}) is given by
\begin{align}
{J}(\omega) =\ & \frac{{{ \varepsilon}}}{{16{\pi ^3}{\varepsilon _0}\hbar }}(  {{\sin }^2}{\theta } ) \nonumber\\&\times  \int {\frac{{{d^3}\vec p}}{{({\omega _e} + {B}{{| {\vec p} |}^2})}}} \delta (\omega  - {\omega _e} - {B}{| {\vec p} |^2}) \nonumber\\
 =\ & \frac{{{\varepsilon }}}{{4{\pi ^2}{\varepsilon _0}\hbar }}(  {{{\sin }^2}{\theta }} ) \nonumber\\&\times \int_0^\infty  {\frac{{{p^2}dp}}{{({\omega _e} + {B}{p^2})}}} \delta (\omega  - {\omega _e} - {B}{p^2}),
\label{TC2}
\end{align}
or for the continuum case
\begin{align}
{J}(\omega ) =\ & \frac{{{\varepsilon}}}{{4{\pi ^2}{\varepsilon _0}\hbar }}(  {{{\sin }^2}{\theta }} )\nonumber\\
&  \times \int_0^\infty  {\frac{{{p^2}dq}}{{({\omega _e} + {B}{p^2})}}\delta (\omega  - {\omega _e} - {B}{p^2})},
\label{Jw}
\end{align}
which leads to
\begin{align}
{J}(\omega ) = \frac{{{\Gamma}{B^{3/2}}}}{\pi }\sum\limits_m {\frac{{p_m^2(\omega )}}{\omega }} \left| {\frac{{Bd{p^2}}}{{dp}}} \right|_{p = {p_m}(\omega )}^{ - 1} ,
\label{Jw1}
\end{align}
with $\Gamma = { \varepsilon }  {{{\sin }^2}({\theta  })/(4\pi {\varepsilon _0}\hbar B^{3/2})} $. The corresponding wave numbers $p_m(\omega)$ are denoted as ${\omega  = {\omega _e} + {B}{p^2(\omega)}}$. Substituting these into Eq.~(\ref{Jw1}), we can get the spectral density (\ref{zwjw}) for the anisotropic three-dimensional photonic crystal environment.

\section{Energy spectrum analysis for the whole system}
\label{section C}
Substituting Eq.~(\ref{eq2}) into the Heisenberg equation $\dot {\hat A}(t) =  - i[\hat A(t),{{\hat H}_H}(t)]$ yields
\begin{eqnarray}
\begin{aligned}
\begin{array}{l}
\dot {\hat a }= - i{\Delta}\hat a + \sum\limits_k {g_k^*\hat b_k^\dag }-i F(t), \
{\dot {\hat b}}_k^\dag = i{\omega _k}\hat b_k^\dag  + {g_k}\hat a,
\end{array}
\label{eqvv38}
\end{aligned}
\end{eqnarray}
where ${{\hat H}_H}(t) = {U^\dag }(t){\hat H}(t)U(t)$ and ${U}(t) = \mathcal {T}{e^{ - i\int_{{0}}^t {{\hat H}({t_1})d{t_1}} }}$ with the operator $\mathcal {T}$ describes the chronological time ordering. It orders any product of operators such that the time argument increases from right to left. From Eq.~(\ref{eqvv38}), the $N \times N$ matrix take the following form:
\begin{eqnarray}
\begin{aligned}
\frac{d}{{dt}}\!\left( {\begin{array}{*{20}{c}}
{\hat a}\\
{\hat b_1^\dag }\\
 \vdots \\
{\hat b_N^\dag }
\end{array}} \right) \!\!=\!\! {\rm{ }}\left( {\begin{array}{*{20}{c}}
{ - i\Delta }&{g_1^*}& \cdots &{g_N^*}\\
{{g_1}}&{i{\omega _1}}& \cdots &0\\
 \vdots & \vdots & \ddots & \vdots \\
{{g_N}}&0& \cdots &{i{\omega _N}}
\end{array}} \right)\!\!\!\left( {\begin{array}{*{20}{c}}
{\hat a}\\
{\hat b_1^\dag }\\
 \vdots \\
{\hat b_N^\dag }
\end{array}} \right) \!-\! i \left( {\begin{array}{*{20}{c}}
F(t)\\
0\\
 \vdots \\
0
\end{array}} \right)
\nonumber
\end{aligned}
\end{eqnarray}
or
\begin{eqnarray}
\begin{aligned}
| {\dot \Phi} \rangle = - i \vartheta \left| \Phi  \right\rangle  -i F(t) \left| \varphi  \right\rangle,
\end{aligned}
\end{eqnarray}
where $\left| \Phi  \right\rangle  = {( {\hat a,\hat b_1^\dag , \cdots ,\hat b_N^\dag } )^T}$, $\left| \Phi  \right\rangle  = {\left( {1,0, \cdots ,0} \right)^T}$, and
\begin{align}
\vartheta = \left( {\begin{array}{*{20}{c}}
{ - i\Delta }&{g_1^*}& \cdots &{g_N^*}\\
{{g_1}}&{i{\omega _1}}& \cdots &0\\
 \vdots & \vdots & \ddots & \vdots \\
{{g_N}}&0& \cdots &{i{\omega _N}}
\end{array}} \right).
\label{C4}
\end{align}
We show that $\vartheta$ in Eq.~(\ref{C4}) denotes the effective non-Hermitian Hamiltonian, whose eigenequation defining the state vector $\left| \psi  \right\rangle  = {(a,{a_1},{a_2} \cdots {a_N})^T}$ reads $\vartheta \left| \psi  \right\rangle  = E_{\rm{BS}}\left| \psi  \right\rangle$, which leads to
\vspace{-1em}
\begin{eqnarray}
a{\Delta} + i\sum\limits_j {g_j^*} {a_j} = E_{\rm{BS}}a,\label{eqvvv42}
\end{eqnarray}
and
\begin{align}
{i{g_k}a - {\omega _k}{a_k} = {a_k}E_{\rm{BS}}},
\label{eqvv42}
\end{align}
by solving Eq.~(\ref{eqvv42}), we get
\begin{eqnarray}
\begin{aligned}
{a_k} = \frac{{i{g_k}a}}{{{\omega _k} + E_{\rm{BS}}}}.
\label{eqvv43}
\end{aligned}
\end{eqnarray}Substituting Eq.~(\ref{eqvv43}) into Eq.~(\ref{eqvvv42}) yields
\begin{eqnarray}
\begin{aligned}
a{\Delta} + i\sum\limits_k {g_k^*} \frac{{i{g_k}a}}{{{\omega _k} + E_{\rm{BS}}}} = E_{\rm{BS}}a.
\label{eqvv44}
\end{aligned}
\end{eqnarray}
Finally, we obtain
\begin{eqnarray}
\begin{aligned}
\Delta  - \sum\nolimits_k {{{|{g_k}{|^2}} \mathord{\left/
 {\vphantom {{|{g_k}{|^2}} {\left( {{\omega _k} + {E_{{\rm{BS}}}}} \right)}}} \right.
 \kern-\nulldelimiterspace} {\left( {{\omega _k} + {E_{{\rm{BS}}}}} \right)}}}  = {E_{\rm{BS}}}
\label{eqvv45}
\end{aligned}
\end{eqnarray}which is consistent with Eq.~(\ref{eqvv10}).

\bibliography{reference}

\end{document}